\documentstyle[11pt,titlepage]{article}
\input epsf
\newcommand{\case}{\left\{\begin{array}{ll}}
\newcommand{\stopcase}{\end{array}\right.}

\newcommand{\proof}{\noindent{\bf Proof: }}
\newcommand{\qed}{\quad\rule{1.5mm}{1.5mm}}

\newcommand{\ds}{\displaystyle}

\newcommand{\la}{\lambda}
\newcommand{\be}{\beta}
\newcommand{\al}{\alpha}
\newcommand{\ep}{\epsilon}
\newcommand{\ga}{\gamma}
\newcommand{\del}{\delta}
\newcommand{\th}{\theta}

\newcommand{\sig}{\sigma}

\newcommand{\Ga}{\Gamma}
\newcommand{\La}{\Lambda}

\newcommand{\om}{\omega}

\newcommand{\R}{{\bf R}}

\newcommand{\M}{{\cal M}}
\newcommand{\Mo}{\M_0}
\newcommand{\Me}{\M_\ep}
\newcommand{\Mx}{\M_x}
\newcommand{\My}{\M_y}

\newcommand{\sgn}{\mbox{sgn}\,}

\newcommand{\co}{{\cal C}_0}
\newcommand{\cmo}{{\cal C}^-_0}
\newcommand{\cxo}{{\cal C}_{x_0}}
\newcommand{\cpi}{{\cal C}_{+\infty}}
\newcommand{\cmi}{{\cal C}_{-\infty}}

\newcommand{\cpu}{{\cal C}^+_u}
\newcommand{\cps}{{\cal C}^+_s}
\newcommand{\cmu}{{\cal C}^-_u}
\newcommand{\cms}{{\cal C}^-_s}

\newcommand{\wu}{W^u}
\newcommand{\ws}{W^s}
\newcommand{\wsme}{\ws(\Me)}
\newcommand{\wume}{\wu(\Me)}
\newcommand{\wuo}{W^u(0)}
\newcommand{\wso}{W^s(0)}

\newcommand{\wsr}{\ws(\Me)}

\newcommand{\wcsr}{\wsme}
\newcommand{\wul}{\wu(0)}
\newcommand{\wcul}{\wuo}
\newcommand{\sy}{{\cal S}^y}
\newcommand{\sz}{{\cal S}^z}
\newcommand{\uy}{{\cal U}^y}
\newcommand{\uz}{{\cal U}^z}
\newcommand{\ux}{{\cal U}^x}
\newcommand{\sx}{{\cal S}^x}
\newcommand{\wsl}{\ws(0)}
\newcommand{\bmo}{B^-_0}
\newcommand{\bpo}{B^+_0}
\newcommand{\bmos}{B^-_{0,s}}
\newcommand{\bpos}{B^+_{0,s}}
\newcommand{\bmxo}{B^-_{x_0}}
\newcommand{\bpxo}{B^+_{x_0}}
\newcommand{\bxo}{B_{x_0}}
\newcommand{\plu}{p^u_0}
\newcommand{\pls}{p^s_0}

\newcommand{\tpo}{t^+_0}
\newcommand{\tmo}{t^-_0}
\newcommand{\txo}{t_{x_0}}
\newcommand{\tpi}{t_{+\infty}}
\newcommand{\tmi}{t_{-\infty}}

\newcommand{\sech}{\mbox{sech}}

\title{Bifurcating bright and dark solitary waves of \\
       the nearly nonlinear cubic-quintic Schr\"odinger
       equation}
       
\author{Todd Kapitula\thanks{Current address: Department of 
        Mathematics and Statistics, University of New Mexico,
        Albuquerque, NM 87131, USA}
   \\ Department of Mathematics
   \\ Virginia Tech
   \\ Blacksburg, VA 24061-0123
   \\ USA}

\date{\today }

\newtheorem{theorem}{Theorem}[section]
\newtheorem{lemma}[theorem]{Lemma}

\newtheorem{prop}[theorem]{Proposition}
\newtheorem{cor}[theorem]{Corollary}

\newtheorem{remark}[theorem]{Remark}

\begin{document}
\maketitle

\noindent ABSTRACT. The existence of bright and dark
multi-bump solitary waves for Ginzburg-Landau type perturbations of 
the cubic-quintic Schr\"odinger equation is considered.
The waves in question are not perturbations of known analytic
solitary waves, but instead arise as a bifurcation from
a heteroclinic cycle in a three-dimensional ODE phase space.
Using geometric singular perturbation techniques, regions in 
parameter space for which 1-bump bright
and dark solitary waves will bifurcate are identified.
The existence of $N$-bump dark solitary waves ($N\ge1$) is shown via an
application of the Exchange Lemma with Exponentially Small Error.
$N$-bump bright solitary waves are shown to
exist as a consequence of the work of Kapitula and Maier-Paape
\cite{kapitula:sdo96}.

\section{Introduction}
\setcounter{equation}{0}

The nonlinear cubic-quintic Schr\"odinger equation is given by
\begin{equation}\label{eqnlcqs}
iA_t=A_{xx}+|A|^2A+\al|A|^4A,
\end{equation}
where $A$ is a complex-valued function of the variables $(x,t)\in
\R\times\R^+$.
When $\al=0$, the equation becomes the focusing
cubic nonlinear Schr\"odinger equation, and is used to
describe the propagation of the envelope of a light pulse in an optical
fiber which has a Kerr-type nonlinear refractive index.
For short pulses and high input peak pulse power the 
refractive index cannot be described by a Kerr-type nonlinearity, as
the index is then influenced by higher-order nonlinearities.
In materials with high nonlinear coefficients, such as 
semiconductors, semiconductor-doped glasses, and organic polymers, 
the saturation of the nonlinear 
refractive-index change is no longer neglible at moderately 
high intensities and should be taken into account (\cite{gatz:spi91}).
Equation (\ref{eqnlcqs}) is the correct model 
to describe the  propagation of the envelope of a light pulse 
in dispersive materials with either a saturable or higher-order
refraction index (\cite{gatz:spi91}, \cite{gatz:sca92}).

Equation (\ref{eqnlcqs}) cannot really be thought of as a 
small perturbation of the cubic nonlinear Schr\"odinger equation,
as it has been shown that a physically realistic value for the
parameter $\al$ is $|\al|\sim 0.1$ (\cite{herrmann:bbs92}).
It turns out that the most physically interesting behavior
occurs when the nonlinearity is saturating, i.e., $\al<0$,
so for the rest of this paper it will be assumed that $\al\sim-0.1$
(\cite{angelis:stp94}, \cite{gatz:spi91}, \cite{gatz:spa92},
\cite{herrmann:bbs92}, \cite{sombra:bpc92}).
An optical fiber which satisfies this condition can be
constructed, for example, by doping with two appropriate materials
(\cite{angelis:stp94}).

One of the more physically interesting phenomena
associated with the double-doped optical fiber is
the existence of bright solitary wave solutions 
($|A(x)|\to0$ as $|x|\to\infty$) in which the peak amplitude becomes 
a two-valued function of the pulse duration.
These solutions have two
different peak powers, and were proven to be stable 
(\cite{boling:oso95}, 
\cite{gatz:spi91}, \cite{grillakis:sto87}, \cite{herrmann:bbs92}).
The solution with the lower peak power corresponds to a perturbation
of the one that exists for the cubic nonlinear Schr\"odinger
equation, while the one with the larger peak power is due
to the saturating nonlinearity.

Equation (\ref{eqnlcqs}) describes an idealized fiber; therefore,
it is natural to consider the perturbative PDE
\begin{equation}\label{eqperturbnlcqs}
iA_t=(1+i\ep a_1)A_{xx}+i\ep\sig A+(1+i\ep d_1)|A|^2A+
    (\al+i\ep d_2)|A|^4A,
\end{equation}
where $0<\ep\ll1$ and the other parameters are real and of $O(1)$.
The parameter $a_1$ describes spectral filtering,
$\sig$ describes the linear gain or loss due to the fiber, and
$d_1$ and $d_2$ describe the nonlinear gain or loss due to
the fiber.
So that (\ref{eqperturbnlcqs}) is a well-defined PDE for $\ep>0$,
it will henceforth be assumed that $a_1>0$.

Solitary wave solutions to (\ref{eqperturbnlcqs}) are found by
setting
\begin{equation}\label{eqsolutionansatz}
A(x,t)=A(x)e^{i\mu t},
\end{equation}
and then finding homoclinic solutions for the ODE
\begin{equation}\label{eqperturbnlcqsodetemp}
(1+i\ep a_1)A''+(\mu+i\ep\sig)A+(1+i\ep d_1)|A|^2A+
    (\al+i\ep d_2)|A|^4A=0,
\end{equation}
where $'=d/dx$.
Multiplying the above by $1-i\ep a_1$ and setting 
\[
x=(1+\ep^2a_1^2)\hat{x}
\]
yields the equivalent ODE
\begin{equation}\label{eqperturbnlcqsodetemp2}
\begin{array}{l}
A''+((\mu+\ep^2a_1\sig)+i\ep(\sig-a_1\mu))A+
   ((1+\ep^2a_1d_1)+i\ep(d_1-a_1))|A|^2A \\
\quad\quad\quad+((\al+\ep^2a_1d_2)+i\ep(d_2-a_1\al))|A|^4A=0,
\end{array}
\end{equation}
where now $'=d/d\hat{x}$.

To simplify matters, all the $O(\ep^2)$ terms in the
above equation will be dropped.
This step is taken only so that the notational complexity is
made as simple as possible, and can clearly be done without
any loss of generality, as for $\ep$ small these terms are
negligible.
Upon dropping these terms, the ODE that will be studied can
finally be written as
\begin{equation}\label{eqperturbnlcqsode}
A''+(\mu+i\ep(\sig-a_1\mu))A+
   (1+i\ep(d_1-a_1))|A|^2A+
   (\al+i\ep(d_2-a_1\al))|A|^4A=0.
\end{equation}

Equation (\ref{eqperturbnlcqsode}) has been extensively studied by
many authors (\cite{de:91}, \cite{duan:fdw95}, \cite{jkp:90}, 
\cite{kapitula:sho95},
\cite{mn:90}, \cite{marcq:eso94}, \cite{saarloos:fps92}).
These papers have been concerned with finding various types of 
solutions, including fronts (kinks), bright solitary waves, and
dark solitary waves.
The methods employed have been both geometric (\cite{de:91}, 
\cite{duan:fdw95}, \cite{jkp:90}, \cite{kapitula:sho95}) and analytic
(\cite{mn:90}, \cite{marcq:eso94}, \cite{saarloos:fps92}).

Bright solitary waves exist when there are solutions to 
(\ref{eqperturbnlcqsode}) which are homoclinic to $|A|=0$.
When $\ep=0$ with
\begin{equation}\label{eqmurangebright}
\frac1{4\al}<\mu<\frac3{16\al},
\end{equation}
one can find a bright solitary wave with a peak power larger than
that associated with the cubic 
nonlinear Schr\"odinger equation.
When $\ep=0$ and $\mu$ is in the range
\begin{equation}\label{eqmurangedark}
\frac3{16\al}<\mu<0
\end{equation}
there exists dark solitary wave solutions.
The dark solitary waves are solutions to (\ref{eqperturbnlcqsode}) which
have the property that $|A(x)|\to A_0\neq0$ as $|x|\to\infty$.

It can be hypothesized that for the perturbative 
PDE (\ref{eqperturbnlcqs}) there
may be a competition between the bright and dark solitary waves for
$\mu$ sufficiently close to the critical value
\begin{equation}\label{eqdefmustar}
\mu^*=\frac3{16\al}.
\end{equation}
It may be further hypothesized that for $\mu$ sufficiently near $\mu^*$ 
it may be possible to construct novel solutions by gluing together 
the bright and dark solitary waves in some way.
The goal of this paper is to explore this possibility.

Doelman \cite{doelman:bth95} has recently considered the problem
of finding $N$-circuit solutions.
These $N$-circuits are constructed by piecing together $N$ copies
of a dark solitary wave, and exist as a solution to 
(\ref{eqperturbnlcqsode}) for $\ep$ sufficiently small and the
parameters in a certain domain in parameter space.
Recent work by De Bouard \cite{bouard:ios95} has shown that the 
dark solitary waves are an unstable solution to (\ref{eqnlcqs}),
due to the presence of a real positive eigenvalue for the operator
obtained by linearizing about the wave.
By a result of Alexander and Jones \cite{alexander:esa94},
since the original solitary wave has an unstable eigenvalue it
is expected that the $N$-circuit solution will have at least
$N$ unstable eigenvalues.

It was previously stated that the bright solitary wave is 
a stable solution to (\ref{eqnlcqs}).
However, recent numerical work by Soto-Crespo et al \cite{crespo:sot96}
shows that this wave
becomes an unstable solution to (\ref{eqperturbnlcqs}) for
$\ep$ nonzero.
The numerics suggest that this instability arises from the presence of a 
real eigenvalue
for the linearized problem 
bifurcating out of the origin and into the right-half of the complex
plane.

Due to these results, when attempting to construct stable solitary
waves to (\ref{eqperturbnlcqs}) one would hope to avoid using 
either the bright or dark waves which exist for (\ref{eqnlcqs}).
This may by possible by considering the situation when 
$\mu=\mu^*$.
When $\ep=0$ and $\mu$ is equal to this critical parameter, there exists
a front solution (kink) to (\ref{eqperturbnlcqsode}) which has the 
asymptotic behavior
\[
|A(x)|\to\left\{\begin{array}{ll}
       0,\quad&x\to-\infty \\
       A_0,\quad&x\to\infty.
       	       \end{array}\right.
\]
Furthermore, since (\ref{eqperturbnlcqsode}) is invariant under
$x\to-x$, there also exists a solution which satisfies
\[
|A(x)|\to\left\{\begin{array}{ll}
       A_0,\quad&x\to-\infty \\
       0,\quad&x\to\infty.
         	\end{array}\right.
\]
Thus, in the ODE phase space there exists a heteroclinic loop,
from which solitary waves may bifurcate as $\ep$ is made
nonzero.
By following the proof in De Bouard \cite{bouard:ios95} it can
be conjectured that for the kink solution the linearized
operator possesses no unstable eigenvalues.
Therefore, it may be possible that any solitary waves 
bifurcating out of the heteroclinic loop may also not have
any unstable eigenvalues.

Based upon the above discussion, it will be of interest to study
the ODE (\ref{eqperturbnlcqsode}) when $\mu$ is (at least) $O(\ep)$ 
close to $\mu^*$.
In particular, it will be of interest to study the dynamics
of the ODE near the heteroclinic cycle for $\ep$ nonzero.
Towards this end, it will first be shown that the cycle
persists for $\ep$ nonzero for $\mu=\mu(\ep)$, with
$\mu(\ep)\to\mu^*$ as $\ep\to0$.
After then fixing $\ep$ the dynamics will be studied as
$\mu$ bifurcates from $\mu(\ep)$.
It is important to note that $\mu$ is a natural bifurcation
parameter in this problem, as it is a free parameter (recall
equation (\ref{eqsolutionansatz})).

This paper will be devoted to proving the following theorems.
Set 
\[
A_0^2=-\frac{3}{4\al},
\]
and define
\[
\begin{array}{lll}
\tilde{d}_1&=&d_1-a_1 \\
\tilde{d}_2&=&d_2-\al a_1 \\
\tilde{\sig}&=&-(\sig+A_0^2d_1+A_0^4d_2)/\ep.
\end{array}
\]
Assume that $\tilde{\sig}=O(1)$ for all $\ep>0$.

Before the main theorems can be stated, a preliminary lemma is
needed regarding the persistence of the kink solitary wave
for $\ep\neq0$.

\begin{lemma}\label{lemgsw} For $0\le\ep\ll1$ there exists
a $\mu(\ep)$, with $|\mu(\ep)-\mu^*|=O(\ep^2)$, such that
a kink solitary wave exists.
\end{lemma}

\begin{remark} Since (\ref{eqperturbnlcqsode}) is
invariant under the transformation $x\to-x$, the above lemma
guarantees the existence of a heteroclinic cycle in the 
ODE phase space for $\ep$ nonzero.
\end{remark}

\begin{theorem}\label{thmbsw} Let $0<\ep\ll1$, and assume that
\[
(\tilde{d}_1+\frac32A_0^2\tilde{d}_2)\tilde{\sig}<0.
\]
There exists a $\mu_h(\ep)<\mu(\ep)$, with $\mu(\ep)-\mu_h(\ep)=
O(e^{-c/\ep})$, such that a bright solitary wave exists.
Furthermore, for each $N\ge2$ there
exists a bi-infinite sequence $\{\mu^N_k\}$ such that when
$\mu=\mu^N_k$ there is an $N$-pulse solution to (\ref{eqperturbnlcqsode}).
The $N$-pulse is even in $x$.
In addition, 
\[
|\mu^N_k-\mu_h(\ep)|=O(e^{-c|k|/\ep})
\]
as $|k|\to\infty$.
\end{theorem}

\begin{remark} An $N$-pulse solution is constructed by
piecing together $N$ copies of a bright solitary wave.
\end{remark}

\begin{remark} The work of Soto-Crespo et al
\cite{crespo:sot96} suggests that the 1-pulse solution
is indeed stable as a solution to (\ref{eqperturbnlcqs}).
\end{remark}

\begin{theorem}\label{thmdsw} Suppose that $0<\ep\ll1$, and that
$0<\mu-\mu(\ep)=O(\ep^n)$ for some $n\ge3$.
Further assume that the parameters satisfy
\[
\begin{array}{l}
(\tilde{d}_1+\frac32A_0^2\tilde{d}_2)\tilde{\sig}<0 \\
(\tilde{d}_1+2A_0^2\tilde{d}_2)\tilde{\sig}<0 \\
(\tilde{d}_1+A_0^2\tilde{d}_2)(\tilde{d}_1+\beta\tilde{d}_2)>0,
\end{array}
\]
where
\[
\beta=A_0^2+\frac14A_0^2(-\ln\frac{\eta}{A_0})^{-1}+O(\eta^2)
\]
for $0<\ep\ll\eta\ll1$.
Then there exists an $N(\ep)>1$, with $N(\ep)\to\infty$ as
$\ep\to0$, such that $N$-circuit solutions exist for $1\le N<N(\ep)$.
\end{theorem}

\begin{remark} The suitable domain in parameter space is shown in
Figure \ref{fig_n_cirpar}.
In this figure the tildes have been dropped, $\be=\al_2/\al_1$,
and $x_0=A_0$.
\end{remark}

The rest of the paper is organized as follows.
In Section 2 the relevant three-dimensional ODE is set up, and the
known solution structure is discussed.
In Section 3 the relevant manifolds in the phase space are 
described.
The bright, dark, and kink solitary waves are formed by
intersecting the appropriate manifolds.
This section is the most technical, and can be skipped
on a first reading.
Sections 4 and 5 discuss the intersection of these manifolds.
In Section 6 the final proofs of the main theorems are given.

\vspace{3mm}
\noindent{\bf Acknowledgements:} I would like to thank Alejandro
Aceves for introducing me to the cubic-quintic Schr\"odinger
equation and some of the references pertaining to it.

\section{Setup of equations}
\setcounter{equation}{0}stable bright solitary
waves are possible for $\mu$ near $\mu^*$.
They do not consider the case of dark solitary waves in their
paper, however.

In this section the first order system of ODEs that will be
studied will be derived.
The goal will be to write the system as compactly as possible so
that the notational burden of the succeeding sections will be
minimized.
Define
\begin{equation}\label{eqdeflambdaomega}
\begin{array}{lll}
\la(r)&=&\mu^*+r^2+\al r^4 \\
\om(r)&=&\tilde{d}_1r^2+\tilde{d}_2r^4,
\end{array}
\end{equation}
where
\begin{equation}\label{eqdeftildedi}
\begin{array}{lll}
\tilde{d}_1&=&d_1-a_1 \\
\tilde{d}_2&=&d_2-a_1\al,
\end{array}
\end{equation}
and furthermore set
\begin{equation}\label{eqdeftildesig}
\begin{array}{lll}
\tilde{\sig}&=&-(\sig-a_1\mu) \\
\mu&=&\mu^*-\tilde{\mu}
\end{array}
\end{equation}
($\mu^*$ is defined in (\ref{eqdefmustar})).
After setting 
\[
A(\hat{x})=r(\hat{x})e^{i\int^{\hat{x}}\phi(s)\,ds}
\]
and substituting this expression into (\ref{eqperturbnlcqsode})
one gets the third order system
\begin{equation}\label{eqodetemp}
\begin{array}{lll}
r'&=&ru \\
u'&=&-u^2-\la(r)+\phi^2+\tilde{\mu} \\
\phi'&=&-2u\phi+\ep(\om(r)-\tilde{\sig}).
\end{array}
\end{equation}
In the above equation the variable $u$ is defined by $u=r'/r$.

The above equations will now be simplified notationally.
First, redefine the variables $(\hat{x},r,u,\phi)$ by
\begin{equation}\label{eqdefineodevariables}
(t,x,y,z)=(\hat{x},r,u,\phi).
\end{equation}
Next, remove the tildes from all the variables.
The ODE (\ref{eqodetemp}) then becomes
\begin{equation}\label{eqgloderedtemp}
\begin{array}{lll}
x'&=&xy \\
y'&=&-y^2+z^2-\la(x)+\mu \\
z'&=&-2yz+\ep(\om(x)-\sig),
\end{array}
\end{equation}
where now $'=d/dt$ and $\om(x)=d_1x^2+d_2x^4$.
For the subsequent analysis it will be useful to set
\[
\sig=\om(x_0)+\ep\tilde\sig,
\]
where $x_0$ is the largest positive root of $\la(x)=0\,(x_0^2=-3/4\al$).
Upon substituting the above expression into (\ref{eqgloderedtemp})
and dropping the tilde, one obtains the equation
\begin{equation}\label{eqglodered}
\begin{array}{lll}
x'&=&xy \\
y'&=&-y^2+z^2-\la(x)+\mu \\
z'&=&-2yz+\ep(\om(x)-\om(x_0)-\ep\sig).
\end{array}
\end{equation}

Equation (\ref{eqglodered}) is the one which will be studied in 
the subsequent sections.
However, it is important to notice that the equation has the
symmetry given in the following proposition.
This symmetry will be used extensively in the upcoming
sections.

\begin{prop}\label{propsymmetry} Equation (\ref{eqglodered}) 
remains invariant under
\[
(t,x,y,z)\to(-t,x,-y,-z).
\]
\end{prop}

\begin{figure}[htb]
\epsfysize=8.5in
\epsffile{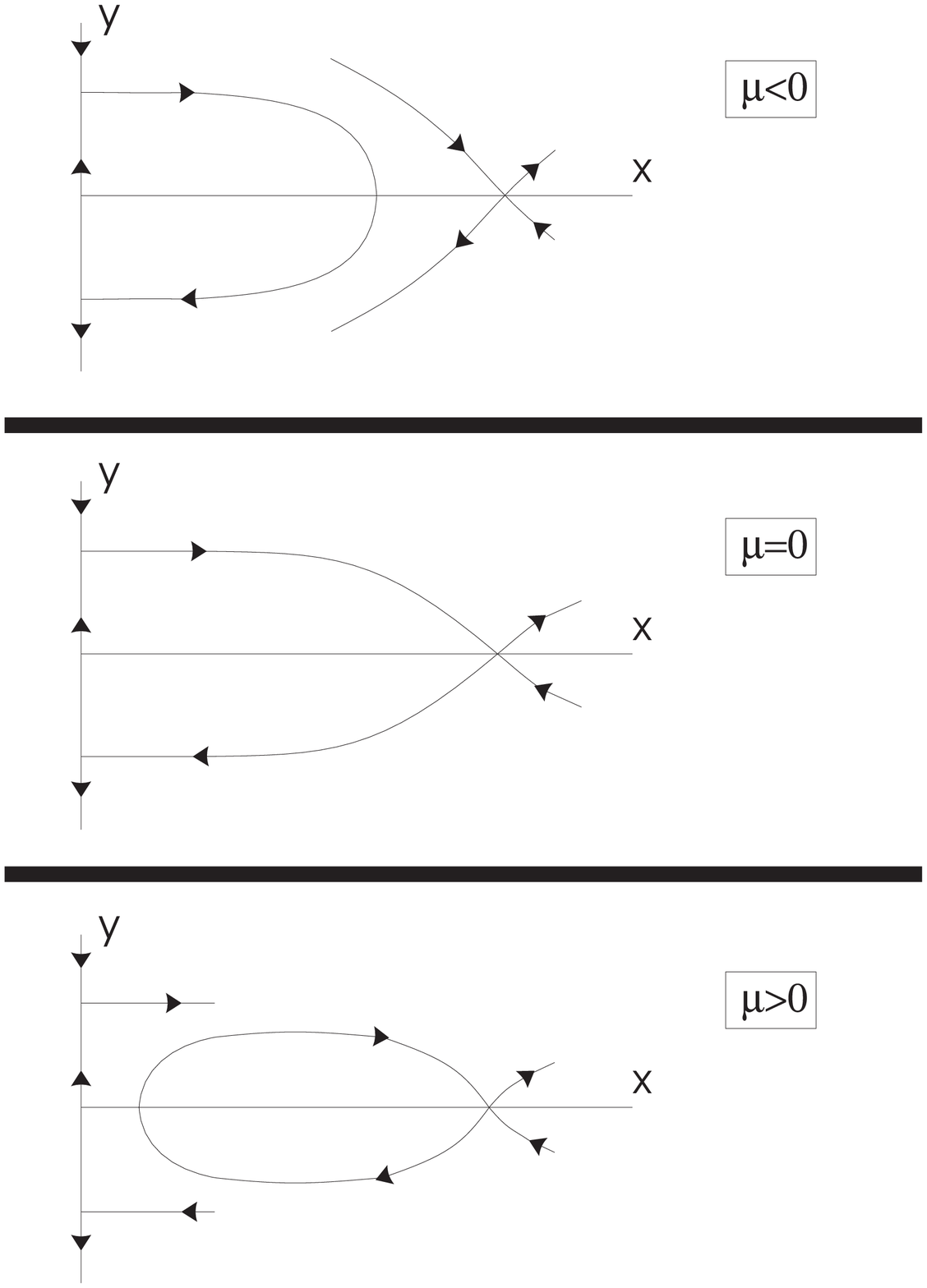}
\caption{Flow on $\{z=0\}$ when $\ep=0$\label{fig_z_0flow}}
\end{figure}

In order to understand the behavior described in the succeeding 
sections, it is first necessary to understand the dynamics of
(\ref{eqglodered}) when $\ep=0$.
In Figure \ref{fig_z_0flow} the flow on the invariant (when $\ep=0$)
plane $\{z=0\}$ is shown for different values of $\mu$.
When $\mu<0$ there exists a heteroclinic connection which connects
the invariant plane $\{x=0\}$ to itself.
This connection corresponds to a bright solitary wave.
For $\mu=0$ there exists a connection between $x=0$ and $x=x_0$.
This solutions corresponds to a kink.
Due to the symmetry described in Proposition \ref{propsymmetry},
there also exists a connection between $x_0$ and zero.
Finally, when $\mu>0$ there exists a solution connecting $x_0$
to itself, which corresponds to a dark solitary wave.
Although it is not shown in the figure, this solution is
actually embedded in the intersection of two two-dimensional
manifolds in the phase space.

In the upcoming sections it will occasionally be useful to have
an analytic expression for the front (kink) which exists when 
$\ep=\mu=0$.
This will be especially true when attempting to gain information
on coefficients used in asymptotic expansions.
Fortunately, due to Marcq et al \cite{marcq:eso94} such an
expression is available.

\begin{prop}\label{propexactwave}  When $\ep=\mu=0$ there exists
a solution $(x(t),y(t),z(t))=(\rho(t),u(t),\phi(t))$ to 
(\ref{eqglodered}) which is given by
\[
\begin{array}{lll}
\rho(t)&=&{\ds\frac{x_0}{\sqrt{2}}(1+\tanh(\frac{x_0}{2}t))^{1/2}} \\
u(t)&=&\rho'(t)/\rho(t) \\
\phi(t)&=&0.
\end{array}
\]
In the above $x_0$ is the largest positive root of $\la(x)=0$ and
is given by
\[
x_0^2=-\frac{3}{4\al}
\]
(recall that $\al\sim-0.1$).
\end{prop}

In the subsequent sections the notation $p\cdot t$ will be
used to represent the trajectory of a point $p=(x,y,z)$ under
the flow generated by (\ref{eqglodered}).
Under this convention, $p\cdot0=p$.
In addition, given a point $p=(x,y,z)$, if it is of interest to
only study the behavior of one of the variables, say $x$, under
the flow, the notation $x\cdot t$ will be used.

\begin{figure}[htb]
\epsfysize=8.5in
\epsffile{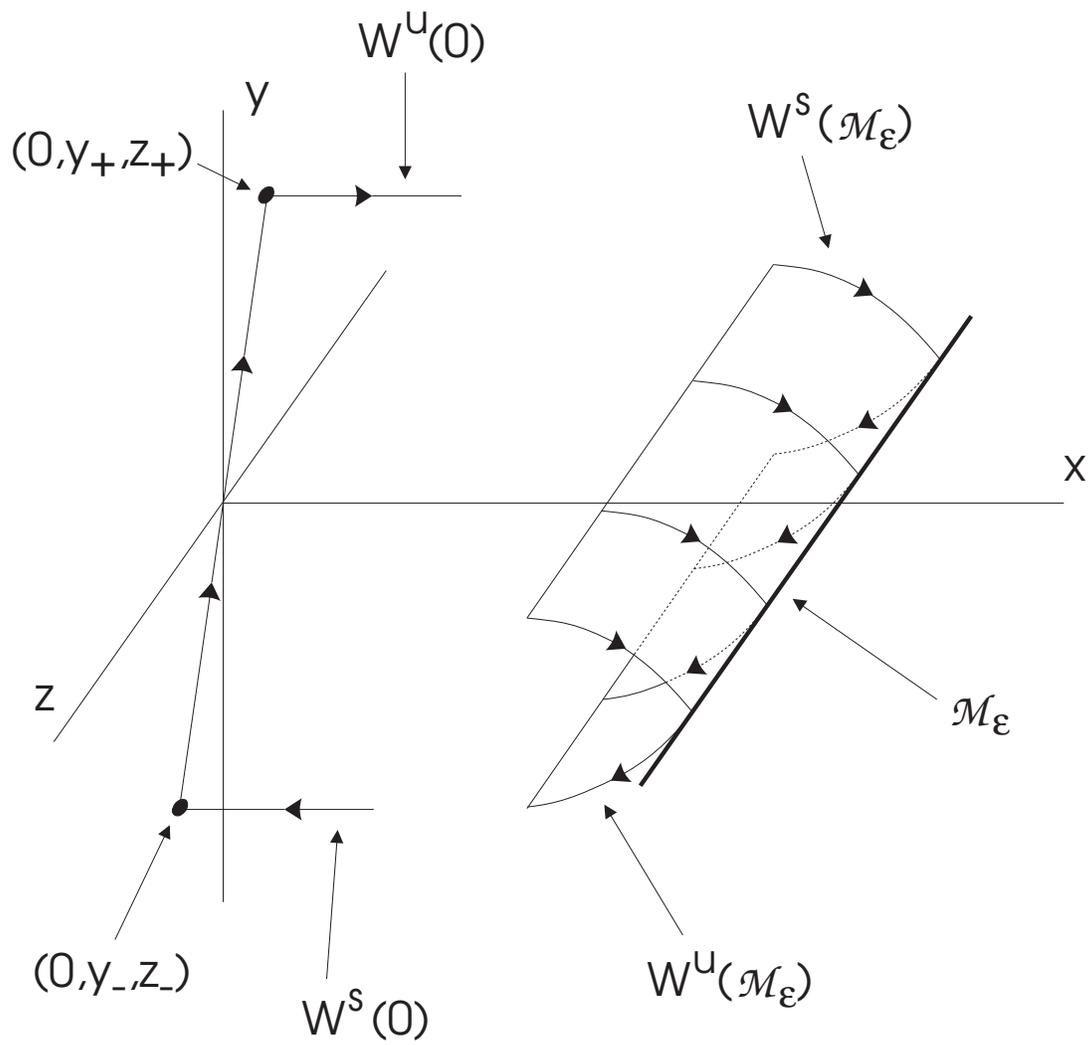}
\caption{The phase space geometry\label{fig_geometry}}
\end{figure}

When discussing the geometric objects in the following sections, the reader
should consult Figure \ref{fig_geometry} to aid in the visualization.


\section{Geometric objects}
\setcounter{equation}{0}

For this section assume that $\mu=\mu(\ep)$, with $\mu(\ep)=O(\ep^2)$.
The validity of this assumption will be verified in Section 4.

\subsection{Flow near $\{x=0\}$}

The goal of this subsection is to characterized the flow near
the invariant plane $\{x=0\}$.
First, there exists a pair of critical points, say
$(0,y_\pm(\ep),z_\pm(\ep))$, which satisfy the algebraic equations
\begin{equation}\label{eqdefypm}
\begin{array}{l}
-y^2+z^2-\la(0)+\mu(\ep)=0 \\
-2yz-\ep(\om(x_0)+\ep\sig)=0.
\end{array}
\end{equation}
It is not difficult to check that
\[
y_\pm(0)=\pm\sqrt{-\la(0)},\quad z_\pm(0)=0;
\]
furthermore, since $\mu(\ep)=O(\ep^2)$,
\begin{equation}\label{eqypmexpand}
\begin{array}{lll}
y_\pm(\ep)&=&{\ds y_\pm(0)+O(\ep^2)} \\
z_\pm(\ep)&=&{\ds-\frac{\om(x_0)}{2y_\pm(0)}\ep+O(\ep^2)}.
\end{array}
\end{equation}
It is also of interest to note that due to the symmetry described
in Proposition \ref{propsymmetry}, 
$y_-=-y_+<0,\,z_-=-z_+$.
Furthermore, upon recalling the expression for $x_0$ given in
Proposition \ref{propexactwave}, it is not difficult to see
that $\la(0)=-x_0^2/4$, and hence $y_\pm(0)=\pm x_0/2$.

For convenience, set
\[
\ga_\pm=y_\pm(0),\quad \al_\pm=-\frac{\om(x_0)}{2y_\pm(0)},
\]
so that the critical points can be said to satisfy the asymptotic
expansion
\[
\begin{array}{lll}
y_\pm(\ep)&=&\ga_\pm+O(\ep^2) \\
z_\pm(\ep)&=&\al_\pm\ep+O(\ep^2).
\end{array}
\]
Linearizing the vector field about the critical points
$(0,y_\pm,z_\pm)$ yields the matrix
\begin{equation}\label{eqdefApm}
A_\pm=\left(\begin{array}{ccc}
   y_\pm & 0 & 0 \\
   0 & -2y_\pm & 2z_\pm \\
   0 & -2z_\pm & -2y_\pm
	    \end{array}\right),
\end{equation}
which has the eigenvalues
\begin{equation}\label{eqdefApmevals}
\begin{array}{lll}
\la^1_\pm&=&\ga_\pm+O(\ep^2) \\
\la^2_\pm&=&-2\ga_\pm+i\al_\pm\ep+O(\ep^2) \\
\la^3_\pm&=&\overline{\la^2_\pm}.
\end{array}
\end{equation}

Thus, the point $(0,y_+,z_+)$ has a one-dimensional unstable manifold,
$\wuo$, coming out of the plane $\{x=0\}$ tangent to the vector
$(1,0,0)$, and a two-dimensional stable manifold embedded in the
plane $\{x=0\}$.
In addition, the point $(0,y_-,z_-)$ has a one-dimensional stable
manifold, $\wso$, coming out of the invariant plane and tangent to
$(1,0,0)$, and a two-dimensional unstable manifold embedded in
the invariant plane.

Given $\del>0$, set
\begin{equation}\label{eqdefco}
\co=\{(x,y,z)\,:\,0\le x\le\del,y=z=0\}.
\end{equation}
Due to the symmetry, this curve will be of paramount importance
in subsequent sections.
In the following proposition the term $\eta$ is assumed to be positive
and small.

\begin{prop}\label{propco} Let $p=(x,y,z)\in\co$ be
such that $x=O(\ep^n)$ for some $n\ge1$.
Then as long as $y_-+\eta\le y\cdot t\le y_+-\eta,\,z\cdot t\neq0$
for $t\neq0$.
Furthermore, $z\cdot(-t)=-z\cdot t$.
\end{prop}

\proof By the nature of the function $\om(x)$ it
is clear that for $x=O(\ep^n),\,\om(x)=O(\ep^{2n})$.
Since $x'=xy,\,x\cdot t=O(x\cdot0)$ as long as 
$y_-+\eta\le y\cdot t\le y_+-\eta$.
Thus, if $x\cdot0=O(\ep^n)$, then $x\cdot t=O(\ep^n)$ for $y\cdot t$
in the prescribed range. 
By the above description of $\om(x)$ for $x$ small, this then
implies that for $x\cdot0=O(\ep^n)$,
\[
z'=-2yz-\ep(\om(x_0)+\ep\sig)+O(\ep^{2n+1}).
\]
If $n\ge1$ it is now clear that trajectories can cross the plane
$\{z=0\}$ at most once for $y_-+\eta\le y\le y_+-\eta$.
This proves the first part of the proposition.
The second part follows immediately from the symmetry inherent
in (\ref{eqglodered}).
\qed

\vspace{3mm}
It is now of interest to determine the nature of the curve $\co$
as the flow carries it by the critical point $(0,y_+,z_+)$.
Using the matrix $A_+$ defined in (\ref{eqdefApm}), the linear flow
near this critical point satisfies
\begin{equation}\label{eqdeflinearcart}
\begin{array}{lll}
x'&=&ax \\
y'&=&-2ay+2\ep bz \\
z'&=&-2\ep bz-2ay,
\end{array}
\end{equation}
where
\[
a=\ga_++O(\ep^2),\quad b=\al_++O(\ep).
\]
Setting $y=r\sin\th,\,z=r\cos\th$ equation (\ref{eqdeflinearcart}) can
be rewritten as
\begin{equation}\label{eqdeflinearpolar}
\begin{array}{lll}
x'&=&ax \\
r'&=&-2ar \\
\th'&=&\ep b,
\end{array}
\end{equation}
which has the solution
\begin{equation}\label{eqdeflinearpolarsoln}
\begin{array}{lll}
x(t)&=&x_0e^{at} \\
r(t)&=&r_0e^{-2at} \\
\th(t)&=&\th_0+\ep bt.
\end{array}
\end{equation}

For each $p\in\co$ define $t_0(p)$ to be such that
\begin{equation}\label{eqdeft0}
{\ds t_0(p)=\{\inf_{t>0}\,:\,p\cdot t\cap\{y_+-r=\eta\}\neq\emptyset\}},
\end{equation}
and set
\begin{equation}\label{eqdefc0t0}
{\ds\co\cdot t_0=\bigcup_{p\in\co}p\cdot t_0(p).}
\end{equation}
After translating the critical point $(0,y_+,z_+)$ to the origin,
the curve $\co\cdot t_0$ can be written parametrically as
\[
\co\cdot t_0=\{(x,r,\th)\,:\,x=x_1(s),r=\eta,\th=-\pi/2+\th_1(s)\},
\]
where $x_1(0)=0,\,x_1'(s)>0,\,x_1(s)\le\eta,$ and $\th_1(s)=O(\ep)$.

It is of interest to determine the nature of $\co\cdot t_0$ as 
the flow forces it to intersect the plane $\{x=\eta\}$.
Given a point in $\co\cdot t_0$, the time of flight, $t_f$,
is defined by $x(t_f)=\eta$.
Sustitution of this expression into (\ref{eqdeflinearpolarsoln})
yields
\begin{equation}\label{eqdeftf}
t_f=\frac1{a}\ln\frac{\eta}{x_1(s)}.
\end{equation}
This expression for $t_f$ further yields that
\begin{equation}\label{eqdefrtf}
\begin{array}{lll}
r(t_f)&=&{\ds\frac1{\eta}x_1^2(s)} \\
\th(t_f)&=&{\ds-\frac{\pi}2+\th_1(s)+\ep\frac{b}{a}\ln\frac{\eta}{x_1(s)}}.
\end{array}
\end{equation}
The above curve is a logarithmic spiral centered upon the point at
which $\wuo$ first intersects $\{x=\eta\}$.
It is important to note that
\begin{equation}\label{eqcospiral}
\begin{array}{lll}
x_1(s)=O(\ep^n)&\Longrightarrow&{\ds\th(t_f)=-\frac{\pi}2+\th_1(s)+
         O(n\ep\ln\frac1{\ep})} \\
x_1(s)=O(e^{-c/\ep})&\Longrightarrow&{\ds\th(t_f)=-\frac{\pi}2+\th_1(s)+
         O(1)}.
\end{array}
\end{equation}
Thus, the spiralling effect is seen only for those points which
are initially exponentially close to the plane $\{x=0\}$.

Define the transverse sections to $\wul$ and $\wsl$ by
\begin{equation}\label{eqdefbpm}
\begin{array}{lll}
\bmo&=&\{(x,y,z)\,:\,x=\eta,|y-y_-|\le\eta,|z-z_-|\le\eta\} \\
\bpo&=&\{(x,y,z)\,:\,x=\eta,|y-y_+|\le\eta,|z-z_+|\le\eta\},
\end{array}
\end{equation}
where, as before, $\eta>0$ is small.
Now define
\begin{equation}\label{eqdefmanint}
\begin{array}{lllll}
\pls&=&\wsl\cap\bmo&=&\{(\eta,y_-+O(\eta),z_-+O(\eta))\} \\
\plu&=&\wul\cap\bpo&=&\{(\eta,y_++O(\eta),z_++O(\eta))\}, 
\end{array}
\end{equation}
where the above represents the first intersection of the manifold with
the set.

Assume that $\del$ is sufficiently small so that for each $p\in\co$
there exists a $t>0$ such that $p\cdot t\cap\bpo\neq\emptyset$.
For each $p\in\co$ define
\begin{equation}\label{eqdeftpo}
{\ds\tpo(p)=\{\inf_{t>0}\,:\,p\cdot t\cap\bpo\neq\emptyset\}},
\end{equation}
and set $\tmo(p)=-\tpo(p)$.
By using a symmetry argument it can be shown that $p\cdot\tmo(p)\in\bmo$.
Finally, define the flow of $\co$ as it intersects the sets
$B^\pm_0$ by
\begin{equation}\label{eqdefcoflow}
\begin{array}{lll}
\co\cdot\tmo&=&{\ds\bigcup_{p\in\co}p\cdot\tmo(p)} \\
\co\cdot\tpo&=&{\ds\bigcup_{p\in\co}p\cdot\tpo(p)}.
\end{array}
\end{equation}
When $\ep=0$ these sets are given by
\begin{equation}\label{eqcolocation}
\begin{array}{lll}
\co\cdot\tmo&=&\{(x,y,z)\,:\,x=\eta,y_-\le y\le y_-+\eta,z=0\} \\
\co\cdot\tpo&=&\{(x,y,z)\,:\,x=\eta,y_+-\eta\le y\le y_+,z=0\}.
\end{array}
\end{equation}

The blow-up sets will be defined next.
This set corresponds to points $p$ such that $|p\cdot t|\to\infty$
in finite time (\cite{kapitula:sdo96}).
As a preliminary, set
\begin{equation}\label{eqdefbpms}
\begin{array}{lll}
\bmos&=&\{(x,y,z)\,:\,0\le x\le\eta,y=y_--\eta,|z-z_-|\le\eta\} \\
\bpos&=&\{(x,y,z)\,:\,0\le x\le\eta,y=y_++\eta,|z-z_+|\le\eta\}.
\end{array}
\end{equation}
Following Kapitula and Maier-Paape \cite{kapitula:sdo96} there exist
sets
\begin{equation}\label{eqdefcpmi}
\begin{array}{lll}
\cmi&=&{\ds\{p\in\bmos\,:\,\exists\,t(p)>0\mbox{ such that }\lim_{t\to t^-(p)}
       p\cdot t=(0,-\infty,0)\}} \\
\cpi&=&{\ds\{p\in\bpos\,:\,\exists\,t(p)<0\mbox{ such that }\lim_{t\to t^+(p)}
       p\cdot t=(0,+\infty,0)\}}.
\end{array}
\end{equation}
The above sets are smooth curves, and symmetry considerations yield
that $(x,y,z)\in\cmi$ implies that $(x,-y,-z)\in\cpi$.
For $p\in\cpi$ set
\begin{equation}\label{eqdeftpi}
\tpi(p)=\{\inf_{t>0}\,:\,p\cdot t\cap\bpo\neq\emptyset\},
\end{equation}
and for $p\in\cmi$ define
\begin{equation}\label{eqdeftmi}
\tmi(p)=\{\sup_{t<0}\,:\,p\cdot t\cap\bmo\neq\emptyset\}.
\end{equation}
Finally, the flow of these sets will be given by
\begin{equation}\label{eqdefcpmiflow}
\begin{array}{lll}
\cmi\cdot\tmi&=&{\ds\bigcup_{p\in\cmi}p\cdot\tmi(p)} \\
\cpi\cdot\tpi&=&{\ds\bigcup_{p\in\cpi}p\cdot\tpi(p)}.
\end{array}
\end{equation}
When $\ep=0$ these sets are given by
\begin{equation}\label{eqcpilocation}
\begin{array}{lll}
\cmi\cdot\tmi&=&\{(x,y,z)\,:\,x=\eta,y_--\eta<y\le y_-,z=0\} \\
\cpi\cdot\tpi&=&\{(x,y,z)\,:\,x=\eta,y_+<y\le y_+-\eta,z=0\}.
\end{array}
\end{equation}

This subsection can now be concluded with the following lemma,
whose conclusion follows from the above discussion.
The fact that $\cmi\cdot\tmi$ and $\cpi\cdot\tpi$ are logarithmic
spirals follows from the same argument as that which led to
the description of $\co\cdot\tmo$ and $\co\cdot\tpo$.

\begin{lemma}\label{lemspirals} Suppose that $\ep\om(x_0)\neq0$.
Then $\co\cdot\tmo$ and $\cmi\cdot\tmi$ are logarithmic spirals
centered on $\pls$, and the curves $\co\cdot\tpo$ and $\cmi\cdot\tpi$
are logarithmic spirals centered on $\plu$.
The spiralling is seen only $O(e^{-c/\ep})$ close to the points
$\pls$ and $\plu$; furthermore, their location is described within
$O(\ep)$ by equations (\ref{eqcolocation}) and (\ref{eqcpilocation}).
\end{lemma}

\begin{remark} By the symmetry described in Proposition \ref{propsymmetry}
 it is necessarily true that
$(\eta,y,z)\in\co\cdot\tmo$ implies that $(\eta,-y,-z)\in\co\cdot\tpo$,
and $(\eta,y,z)\in\cmi\cdot\tmi$ implies that 
$(\eta,-y,-z)\in\cpi\cdot\tpi$.
\end{remark}

\subsection{Flow near $\Me$}

The purpose of this subsection is to describe the flow generated
by (\ref{eqglodered}) near the point $(x_0,0,0)$, where $x_0$
is the largest positive zero of $\la(x)=0$.

When $\ep=0$ there exists a critical manifold $\Mo$ which is given by
\begin{equation}\label{eqm0}
\Mo=\{(x,y,z)\,:\,y=0,\,z^2-\la(x)=0,\,\,4z^2+x\la'(x)<0\}.
\end{equation}
This manifold is normally hyperbolic, and therefore smoothly perturbs
to a manifold $\Me$ for $\ep$ sufficiently small. 
Furthermore, the manifold $\Me$ has a two-dimensional stable
manifold, $\wsme$, and a two-dimensional unstable manifold,
$\wume$, which are smooth perturbations of the center-stable
and center-unstable manifolds which exist when $\ep=0$
(Fenichel \cite{fe:73}, Jones \cite{jones:gsp95}).

As in the previous section, assume that $\mu=\mu(\ep)=O(\ep^2)$.
The critical points for (\ref{eqglodered}) which lie on $\Me$ 
satisfy $y=0$ and
\begin{equation}\label{eqcriticalpts}
\begin{array}{l}
\la(x)-z^2+O(\ep^2)=0 \\
\om(x)-\om(x_0)-\ep\sig=0.
\end{array}
\end{equation}
A Taylor expansion about the point $x_0$ yields that
\begin{equation}\label{eqcritpertom}
x-x_0=\ep\frac1{\om'(x_0)}+O(\ep^2).
\end{equation}
Substituting the expression in (\ref{eqcritpertom}) into the first
equation of (\ref{eqcriticalpts}) and taking a Taylor expansion
yields
\begin{equation}\label{eqcritpertla}
\ep\frac{\la'(x_0)}{\om'(x_0)}-z^2+O(\ep^2)=0,
\end{equation}
from which one gets that
\begin{equation}\label{eqzcritpts}
z^2=\ep\frac{\la'(x_0)}{\om'(x_0)}+O(\ep^2).
\end{equation}
This in turn yields
\begin{equation}
z=\pm\sqrt{\la'(x_0)\frac{\sig}{\om'(x_0)}}\ep^{1/2}+O(\ep).
\end{equation}
Letting $(x^*(\ep),0,\pm z^*(\ep))\in\Me$ represent the critical points,
it is now seen that these points have the asymptotic expansion
\begin{equation}\label{eqdefxszs}
\begin{array}{lll}
x^*(\ep)&=&{\ds x_0+\frac{\sig}{\om'(x_0)}\ep+O(\ep^2)} \\
z^*(\ep)&=&{\ds\sqrt{\la'(x_0)\frac{\sig}{\om'(x_0)}}\ep^{1/2}+O(\ep)}.
\end{array}
\end{equation}
Since $\la'(x_0)<0$, in order for the above expressions to make 
sense the following assumption must hold.

\begin{prop}\label{asssig_1} If the parameters $d_1,d_2$, and $\sig$
are chosen so that
\[
\om'(x_0)\sig<0,
\]
i.e.,
\[
(d_1+2x_0^2d_2)\sig<0,
\]
then there exist critical points on $\Me$ whose $x$ and $z$ components
are given by (\ref{eqdefxszs}).
\end{prop}


The next goal is to get an asymptotic description of $\Me$ 
within $O(\ep)$ of $\{z=0\}$.
A preliminary lemma is first needed.
In the following, let the curve $\cxo$ be given by
\begin{equation}\label{eqdefcxo}
\cxo=\{(x,y,z)\,:\,y=z=0,|x-x_0|\le\nu\}.
\end{equation}

\begin{lemma}\label{lemcxome} There exists an $\ep_0>0$ such
that if $0\le\ep<\ep_0$, then $\cxo\cap\Me\neq\emptyset$.
\end{lemma}

\proof Clearly, $\Me$ intersects the plane $\{z=0\}$ nontrivially.
Set
\[
\{(x_\ep,y_\ep,0)\}=\Me\cap\{z=0\}.
\]
In order to prove the lemma, it must be shown that $y_\ep=0$ for $\ep\neq0$.
By the definition of $\Mo$ and the fact that $\Me$ is a smooth perturbation
of $\Mo$ it is clear that 
\[
{\ds\lim_{\ep\to0}y_\ep=0.}
\]

In a sufficiently small neighborhood of $\Me$ the manifolds $\wsme$ and
$\wume$ satisfy the relation
\[
\wsme\cap\wume=\Me.
\]
This immediately yields that
\begin{equation}\label{eqxepyep}
\wsme\cap\wume\cap\{z=0\}=\{(x_\ep,y_\ep,0)\}.
\end{equation}
Suppose without loss of generality that
\[
{\ds\lim_{t\to\pm\infty}(x_\ep,y_\ep,0)\cdot t=(x^*,0,\pm z^*)}.
\]
By the symmetry of the ODE, if $(x_\ep,y_\ep,0)\cdot t$ is a 
solution, then so is $(x_\ep,-y_\ep,0)\cdot(-t)$.
This immediately yields that
\[
{\ds\lim_{t\to\pm\infty}(x_\ep,-y_\ep,0)\cdot t=(x^*,0,\mp z^*)},
\]
so that
\[
\{(x_\ep,-y_\ep,0)\}\subset\wsme\cap\wume\cap\{z=0\}.
\]
But equation (\ref{eqxepyep}) then gives
\[
(x_\ep,y_\ep,0)=(x_\ep,-y_\ep,0)
\]
so that $y_\ep=0$.
\qed

\vspace{3mm}
For the rest of this subsection set $\zeta=|z|+\ep$.
The slow manifold $\Me$ is given by the graph
\begin{equation}\label{eqdefme}
\Me=\{(x,y,z,e)\,:\,x=\Mx(z,\ep),y=\My(z,\ep)\}.
\end{equation}
By the definition of $\Mo$ the function $\My$ satisfies
the relation $\My(z,0)=0$.
Furthermore, the conclusion of Lemma \ref{lemcxome} yields that
$\My(0,\ep)=0$.
Thus, a Taylor expansion of $\My$ gives
\begin{equation}\label{eqdefmy}
\My(z,\ep)=\ep z\tilde\My(z,e),
\end{equation}
where the function $\tilde\My$ is uniformly bounded for $\zeta$
sufficiently small.
The following lemma gives a description of the functions $\Mx$ and
$\tilde\My$.

\begin{lemma}\label{lemdescribeme} The functions $\Mx$ and $\My$ 
comprising the graph of $\Me$ have the Taylor expansion
\[
\begin{array}{lll}
\Mx(z,\ep)&=&{\ds x_0+\frac1{\la'(x_0)}z^2+\tilde{x}\ep^2+O(\zeta^3)}\\
\My(z,\ep)&=&{\ds\ep z(-\frac{2\sig}{x_0\la'(x_0)}\ep+O(\zeta^2))},
\end{array}
\]
where $\tilde x$ is an $O(1)$ constant.
\end{lemma}

\proof The basic idea of the proof is to write out a Taylor expansion
for both $\Mx$ and $\My$ and to use the fact that the manifold $\Me$
is invariant under the flow.
After using (\ref{eqdefmy}) the functions can be expanded as
\[
\begin{array}{lll}
\Mx(z,\ep)&=&x_0+x_1z+x_2\ep+x_3z^2+x_4\ep z+x_5\ep^2+O(\zeta^3) \\
\My(z,\ep)&=&\ep z(y_0+y_1z+y_2\ep+O(\zeta^2)),
\end{array}
\]
where $x_0$ is such that $\la(x_0)=0$.
In order to prove the lemma, it must be shown that
\[
x_1=x_2=x_4=y_0=y_1=0
\]
and 
\[
x_3=\frac1{\la'(x_0)},\,y_2=-\frac{2\sig x_3}{x_0}.
\]

When $\ep=0$ the function $\Mx$ satisfies the relation
\[
z^2-\la(\Mx(z,0))=0.
\]
Taking a Taylor expansion of $\la(x)$ about $x_0$ and equating
coefficients quickly yields the desired relation for $x_1$ and
$x_3$.

In order to show that $x_2=0$, consider the following 
argument.
The flow on $\Me$ satisfies
\begin{equation}\label{eqmegeneralflow}
z'=-2\My(z,\ep)z+\ep(\om(\Mx(z,\ep))-\om(x_0)-\ep\sig),
\end{equation}
so that when $z=0$
\[
z'=\ep((x_2\om'(x_0)-\sig)\ep+O(\ep^2)).
\]
The above expression is arrived at upon taking a Taylor expansion
of $\om(x)$ about $x_0$ and using the expansion for $\Mx(0,\ep)$.
Note that $z'=O(\ep^2)$.
Now, the fact that on $\Me,\,y=\My(z,\ep)$ implies that
\[
y'=z'\partial_z\My(z,\ep).
\]
Using the Taylor expansion for $\My$ and the fact that $z'=O(\ep^2)$
yields that when $z=0,\,y'=O(\ep^3)$.
But when $z=0$,
\[
y'=-\My^2(0,\ep)-\la(\Mx(0,\ep))+\mu(\ep).
\]
Since $\mu(\ep)=O(\ep^2)$, after taking a Taylor expansion of 
$\la(x)$ about $x_0$ one gets that
\[
y'=-x_2\la'(x_0)\ep+O(\ep^2).
\]
Thus, the fact that $y'=O(\ep^3)$ necessarily gives that $x_2=0$.

Using the above results and equation (\ref{eqmegeneralflow}) gives
that the equation for $z$ on $\Me$ is
\[
z'=-\sig\ep^2+(x_3\om'(x_0)-2y_0)\ep z^2+x_4\om'(x_0)\ep^2z
  +x_5\om'(x_0)\ep^3+O(\zeta^4).
\]
Since $x'=xy$, on $\Me$
\begin{equation}\label{eqmxprime}
\Mx'=\Mx\My.
\end{equation}
But
\[
\begin{array}{lll}
\Mx'&=&z'\partial_z\Mx \\
{ } &=&-2\sig x_3\ep^2z-\sig x_4\ep^3+O(\zeta^4)
\end{array}
\]
and
\[
\Mx\My=x_0y_0\ep z+x_0y_1\ep z^2+x_0y_2\ep^2z+O(\zeta^4),
\]
so that (\ref{eqmxprime}) becomes
\[
-2\sig x_3\ep^2z-\sig x_4\ep^3+O(\zeta^4)=
    x_0y_0\ep z+x_0y_1\ep z^2+x_0y_2\ep^2z+O(\zeta^4).
\]
Upon equating coefficients the desired result is reached.
\qed

\vspace{3mm}
\subsection{Flow near $\Me$}

Recall that the flow on $\Me$ is given by (\ref{eqmegeneralflow}).
Using the result of Lemma \ref{lemdescribeme} and Taylor expanding
$\om(x)$ around $x_0$ then yields that on $\Me$ the flow is
given by
\begin{equation}\label{eqmeflowestimate}
{\ds z'=\ep(\frac{\om'(x_0)}{\la'(x_0)}z^2-\sig\ep+O(\ep^2))+O(\zeta^4)}.
\end{equation}
Note that the above equation implies that for $|z|=O(\ep)$,
\begin{equation}\label{eqmeflowzsmall}
z'=-\ep^2\sig+O(\ep^3).
\end{equation}
This estimate will turn out to be crucial for the estimates that follow.

In order to determine the nature of the flow near $\Me$, it will
be desirous to use Fenichel coordinates (\cite{jones:gsp95},
\cite{jones:tif93}, \cite{jones:tim96}).
Before doing so, it will first be beneficial to understand the
matrix arrived at upon linearizing about $\Me$.
Specifically, the manner in which the eigenvalues and eigenvectors
vary as a function of $z$ and $\ep$ must be understood.

Linearizing the flow (\ref{eqglodered})
about a point $(x,y,z)\in\Me$ yields the matrix
\begin{equation}\label{eqmelinear}
A_{\Me}=\left(\begin{array}{rrr}
   y & x & 0 \\
   -\la'(x) & -2y & 2z \\
   \ep\om'(x) & -2z & -2y,
      \end{array}\right)
\end{equation}
which has the characteristic equation
\begin{equation}\label{eqmechareq} 
P(\ga,z,\ep)=(\ga-y)((\ga+2y)^2+4z^2)+x(\la'(x)(\ga+2y)-2\ep\om'(x)z)=0.
\end{equation}
The above equation has two solutions which are $O(1)$, say $\ga^\pm(z,\ep)$,
with $\ga^+=-\ga^-$ and $\ga^-(0,0)=-\sqrt{-x_0\la'(x_0)}$.

In the following calculations the result of Lemma \ref{lemdescribeme}
will always be implicitly used.
It is easy to check that 
\[
P_\ga(\ga^-,0,0)=-2x_0\la'(x_0)
\]
and 
\[
P_z(\ga^-,0,0)=P_\ep(0,0)=0,
\]
so that 
\[
\ga^-_z(0,0)=\ga^-_\ep(0,0)=0.
\]
The above follows immediately from the fact that
\[
\ga^-_z=-P_z/P_\ga,\quad\ga^-_\ep=-P_\ep/P_\ga.
\]

Taking second derivatives gives 
\[\begin{array}{lll}
P_{zz}(\ga^-,0,0)&=&-8\sqrt{-x_0\la'(x_0)} \\
P_{z\ep}(\ga^-,0,0)&=&-2x_0\om'(x_0).
  \end{array}
\]
Since
\[
\ga_{zz}=-P_{zz}/P_\ga,\quad\ga_{z\ep}=-P_{z\ep}/P_\ga,
\]
this immediately gives that
\[\begin{array}{lll}
\ga^-_{zz}(0,0)&=&{\ds\frac4{\sqrt{-x_0\la'(x_0)}}} \\
\ga^-_{z\ep}(0,0)&=&{\ds-\frac{\om'(x_0)}{\la'(x_0)}}.
  \end{array}
\]
The proof of the following proposition is now complete. 

\begin{prop}\label{propevalsme} The $O(1)$ eigenvalues $\ga^\pm$ 
of $A_{\Me}$ satisfy
\[\begin{array}{ll}
1)\quad&\ga^+(z,\ep)=-\ga^-(z,\ep) \\
2)\quad&{\ds\ga^-(z,\ep)=-\sqrt{-x_0\la'(x_0)}+
                  \frac2{\sqrt{-x_0\la'(x_0)}}z^2- 
                  \frac{\om'(x_0)}{\la'(x_0)}\ep z+
                  \tilde{\ga}\ep^2+O(\zeta^3)},
  \end{array}
\]
where $\tilde{\ga}$ is an $O(1)$ constant.
\end{prop}

Although the following information is not necessary at the
moment, it will be useful at a later time.
The eigenvector ${\bf v}_{\ga^-}$ of $A_{\Me}$ associated with
the eigenvalue $\ga^-$ is given by
\begin{equation}\label{eqevector}
{\ds{\bf v}_{\ga^-}=(x,\ga^--y,\frac{-2(\ga^--y)z+\ep x\om'(x)}{\ga^-+2y})^T}.
\end{equation}
Let ${\bf v}_{\ga^-}=(v_1,v_2,v_3)^T$.
After using the expansions given in Lemma \ref{lemdescribeme} and 
Proposition \ref{propevalsme} the following proposition is realized.

\begin{prop}\label{propevectsme} The components of the eigenvector 
${\bf v}_{\ga_-}$ are given by the expansions
\[
\begin{array}{lll}
v_1&=&{\ds x_0+\frac1{\la'(x_0)}z^2+\tilde{x}\ep^2+O(\zeta^3)}\\
v_2&=&{\ds-\sqrt{-x_0\la'(x_0)}+
                  \frac2{\sqrt{-x_0\la'(x_0)}}z^2- 
                  \frac{\om'(x_0)}{\la'(x_0)}\ep z+
                  \tilde{\ga}\ep^2+O(\zeta^3)} \\
v_3&=&{\ds-2z-\frac{x_0\om'(x_0)}{\sqrt{-x_0\la'(x_0)}}\ep+O(\zeta^2)},
\end{array}
\]
where $\tilde{x}$ and $\tilde{\ga}$ are $O(1)$ constants.
\end{prop}

\begin{remark} Due to the symmetry of the ODE, the eigenvector 
associated with $\ga^+$ is such that the second and third 
components are the negative of those given above.
\end{remark}

\begin{remark}\label{remstrongstable}
When the above expressions are evaluated at the critical points
$(x^*(\ep),0,\pm z^*(\ep))$, it is of interest to note
that the third component, $v_3$, satisfies the estimate
\[
v_3=-2(\pm z^*(\ep))+O(\ep).
\]
The above expression is valid because it is known that 
$z^*(\ep)=O(\ep^{1/2})$.
\end{remark}

Recall that for $|z|=O(\ep)$ that the flow on $\Me$ satisfies
\[
z'=-\ep^2\sig+O(\ep^3).
\]
Thus, in Fenichel coordinates the equations near $\Me$ for 
$|z|=O(\ep)$ may be written as
\begin{equation}\label{eqfenichel}
\begin{array}{lll}
a'&=&\La(a,b,z,\ep)a \\
b'&=&\Ga(a,b,z,\ep)b \\
z'&=&\ep^2(-\sig+h(a,b,z,\ep)ab).
\end{array}
\end{equation}
In the above equation the set $a=0$ refers to $\wsme$, the set
$b=0$ refers to $\wume$, and the function $h$ is uniformly 
bounded.
Using the eigenvalue expansion in Proposition \ref{propevalsme}
one can say that for $|z|=O(\ep)$,
\[
\La(0,0,z,\ep)=\sqrt{-x_0\la'(x_0)}+O(\ep^2),\quad
\Ga(0,0,z,\ep)=-\La(0,0,z,\ep).
\]
After taking a Taylor expansion the equations (\ref{eqfenichel})
may then be rewritten as
\begin{equation}\label{eqfenichel2}
\begin{array}{lll}
a'&=&\sqrt{-x_0\la'(x_0)}\,a+O((|a|+|b|+\ep)^2) \\
b'&=&-\sqrt{-x_0\la'(x_0)}\,b+O((|a|+|b|+\ep)^2)\\
z'&=&\ep^2(-\sig+h(a,b,z,\ep)ab).
\end{array}
\end{equation}

Define the set $\bxo$ by
\begin{equation}\label{eqdefbxo}
\bxo=\{(a,b,z)\,:\,|a|+|b|\le\nu,|z|\le O(\ep)\}.
\end{equation}
For $\ep$ and $\nu$ sufficiently small it is clear that one can, without
loss of generality, use the linear equations
\begin{equation}\label{eqfenichellinear1}
\begin{array}{lll}
a'&=&\sqrt{-x_0\la'(x_0)}\,a \\
b'&=&-\sqrt{-x_0\la'(x_0)}\,b \\
z'&=&-\ep^2\sig
\end{array}
\end{equation}
when discussing the flow in $\bxo$.
For convenience, in the rest of this subsection $\ga=\sqrt{-x_0\la'(x_0)}$,
so that (\ref{eqfenichellinear1}) can be rewritten as
\begin{equation}\label{eqfenichellinear}
\begin{array}{lll}
a'&=&\ga a \\
b'&=&-\ga b \\
z'&=&-\ep^2\sig.
\end{array}
\end{equation}

Now that the equations for the flow near $\Me$ have been determined,
it will be of interest to determine the nature of the set $\cxo$
as it exits $\bxo$ under the influence of the flow.
To be precise, for each $p\in\cxo$ define $t_{x_0}$ by
\[
{\ds t_{x_0}(p)=\{\sup_{t<0}\,:\,p\cdot t\in\bpxo\}},
\]
and set
\begin{equation}\label{eqdefcxoflow}
{\ds\cxo\cdot\txo=\bigcup_{p\in\cxo}p\cdot t_{x_0}(p).}
\end{equation}
The set $\cxo\cdot\txo$ will be close to the set $a=0$, i.e.,
close to $\wsme$, as it exits $\bpxo$, where
\begin{equation}\label{eqdefbpxo}
\bpxo=\{(x,y,z)\,:\,x=x_0-\nu,|y-(\sqrt{-\la'(x_0)/x_0})\nu|\le\nu,
        |z|\le\nu\}
\end{equation}
The following lemma gives a determination as to how close.

\begin{lemma}\label{lemimagecxo} Let $p\in\cxo$ be such that 
$\txo(p)=O(1/\ep)$.
The curve $\cxo\cdot\txo$ is $C^1-O(e^{-c/\ep})$ close to
$a=0$ in a neighborhood of $p\cdot\txo(p)$,
where $c>0$ is some constant.
\end{lemma}

\proof In order to prove the lemma the time-reversed flow for
(\ref{eqfenichellinear}) must be considered.
After such a time reversal, the solution to (\ref{eqfenichellinear}) is
given by
\begin{equation}\label{eqfenproof}
\begin{array}{lll}
a(t)&=&a_0e^{-\ga t} \\
b(t)&=&b_0e^{\ga t} \\
z(t)&=&z_0+\ep^2\sig t.
\end{array}
\end{equation}
In Fenichel coordinates the curve $\cxo$ can be parametrically represented 
as $(a_0(s),b_0(s),z_0(s))=(s,s,0)$ with $0\le s\le\nu$.
Substituting this set of initial conditions into (\ref{eqfenproof}) yields 
that the flow of $\cxo$ is governed by
\begin{equation}\label{eqfenproof2}
\begin{array}{lll}
a(t)&=&se^{-\ga t} \\
b(t)&=&se^{\ga t} \\
z(t)&=&\ep^2\sig t.
\end{array}
\end{equation}

A solution to (\ref{eqfenproof2}) leaves $\bxo$ when $b(t)=\nu$.
Solving this equation yields that the time-of-flight, $t_f$, is
given by
\[
t_f=\frac1{\ga}\ln\frac{\nu}{s}.
\]
Substituting $t_f$ into (\ref{eqfenproof2}) yields
\begin{equation}\label{eqfenproof3}
\begin{array}{lll}
a(t_f)&=&{\ds\frac{s^2}{\nu}} \\
z(t_f)&=&{\ds-\ep^2\frac1{\ga}\ln\frac{\nu}{s}}.
\end{array}
\end{equation}

The above equation is a parametric representation of a curve on
the section $\bpxo$.
It can be rewritten as
\[
a(z)=\nu e^{\be z/\ep^2},
\]
where 
\[
\be=2\ga/\sig,\quad\sig z<0.
\]
The conclusion of the lemma is now clear, as when $z=O(\ep)$ with 
$\sig z<0$,
\[
a(z)=O(e^{-c/\ep}),\quad a'(z)=O(\frac{e^{-c/\ep}}{\ep^2}),\quad\qed.
\]

\vspace{3mm}


\section{$\wuo\cap\wsme$ transversely}
\setcounter{equation}{0}

After appending $\mu'=0$ and $\ep'=0$ to (\ref{eqglodered})
one arrives at a three-dimensional manifold $\wcul$ (abusing notation
here) and a
four-dimensional manifold $\wcsr$ in the five-dimensional phase
space.
Since $x'>0$ along the relevant trajectory for $0<x<x_0$, by 
continuity this holds for $\ep$ and $\mu$ sufficiently small.
As such, for $0<x<x_0$ the manifolds can be parameterized in 
the following manner:
\begin{equation}\label{eqdeffolman}
\begin{array}{lll}
\wuo&=&\{(x,y,z)\,:\,y=\uy(x,\mu,\ep),z=\uz(x,\mu,\ep)\} \\
\wsme&=&\{(x,y,z)\,:\,y=\sy(x,z,\mu,\ep),|z|\le\tilde\ep\ll1\}.
\end{array}
\end{equation}
A heteroclinic orbit possibly exists when $\wcul\cap\wcsr$ nontrivially, or 
equivalently when the function
\begin{equation}\label{eqdeffunintersect}
F(x,\mu,\ep)=\sy(x,\uz(x,\mu,\ep),\mu,\ep)-\uy(x,\mu,\ep)
\end{equation}
has a zero.

It is clear that $F(x,0,0)=0$.
If $F_\mu(x,0,0)\neq0$ for some value of $x\in(0,x_0)$, then an application
of the implicit function theorem yields that there exists a smooth
function $\mu=\mu(\ep)$ such that $F(x,\mu(\ep),\ep)=0$.
In other words, the condition that $F_\mu\neq0$ gives that the manifolds
$\wcul$ and $\wcsr$ intersect transversely in $(x,y,z,\mu)$-space, so that 
their intersection persists for small perturbation.
Furthermore, the implicit function theorem yields that
\begin{equation}\label{eqmuprime}
\mu'(0)=-\frac{F_\ep(x,0,0)}{F_\mu(x,0,0)}.
\end{equation}
Thus, it is possible to get asymptotics for the function $\mu(\ep)$ by
thoroughly understanding the manner in which the manifolds intersect.

Differentiating (\ref{eqdeffunintersect}) yields that
\begin{equation}\label{eqdeffmutemp}
F_\mu(x,0,0)=(\sy_\mu-\uy_\mu+\sy_z\uz_\mu)(x,0,0).
\end{equation}
An examination of (\ref{eqglodered}) yields that when $\ep=0$ it
is unchanged under the transformation $z\to-z$; therefore, 
\begin{equation}\label{eqsy=0}
\sy_z(x,0,\mu,0)=0.
\end{equation}
This gives the simplified expression of $F_\mu$ to be
\begin{equation}\label{eqdeffmu}
F_\mu(x,0,0)=\sy_\mu(x,0,0,0)-\uy_\mu(x,0,0).
\end{equation}
In a similiar manner, it is seen that
\begin{equation}\label{eqdeffep}
F_\ep(x,0,0)=\sy_\ep(x,0,0,0)-\uy_\ep(x,0,0).
\end{equation}

Let the heteroclinic orbit which exists for $\mu=\ep=0$ be denoted
by $(\rho(t),u(t),0)$.
Assume that the wave has been translated so that $\rho(0)=x_1$,
where $0<x_1<x_0$ is the other root of $\la(x)=0$. 
When linearizing (\ref{eqglodered}) about this orbit the variational
equations take the form
\begin{equation}\label{eqodevar}
\begin{array}{lll}
\del x'&=&u\del x+\rho\del y \\
\del y'&=&-\la'(\rho)\del x-2u\del y+\del\mu \\
\del z'&=&-2u\del z+(\om(\rho)-\om(x_0))\del\ep \\
\del\mu'&=&0 \\
\del\ep'&=&0.
\end{array}
\end{equation}
Each of the coordinates $\del x_i$ of (\ref{eqodevar}) ($x_i=x,y$, etc.)
is a 1-form.
From these 1-forms exterior products of forms of any degree $k$ can be 
constructed.
The rest of the following discussion will only be concerned
with 2-forms $P_{x_ix_j}=\del x_i\wedge\del x_j$; however, it can
be generalized to any $k$.

Each 2-form associates to a 2-plane $T$ a number that is the
area of the projection of a unit square of $T$ onto the coordinate planes
of the two coordinates specified by $P_{x_ix_j}$.
In order to evaluate $P_{x_ix_j}$, the following rule is applied.
Let $N$ represent a $k$-dimensional manifold, $k\ge2$, and let $T_pN$
represent the tangent space to $N$ at a point $p\in N$.
Let $\{a_i(p)\}$ represent a basis for $T_pN$.
Suppose that $S(p)\subset T_p(n)$ represents a two-dimensional subspace
which is spanned by the vectors $\{a_i(p),a_j(p)\}$ for some
$1\le i,j\le k$.
These vectors may be thought of as rows to a $2\times k$ matrix, $A(p)$.
If $x_i$ and $x_j$ represent two of the $k$ coordinates of $T_pN$, then
$P_{x_ix_j}$ is the determinant of the $2\times2$ submatrix obtained by
looking at the $x_i^{\scriptscriptstyle{\mbox{th}}}$ and 
$x_j^{\scriptscriptstyle{\mbox{th}}}$ columns of  $A(p)$.
Furthermore, the evolution equation for $P_{x_ix_j}$ is given by
the product rule, i.e., 
\[
P'_{x_ix_j}=\del x'_i\wedge\del x_j+\del x_i\wedge\del x'_j.
\]

\begin{lemma}\label{lemuymu} For any $0<x<x_0,\,\uy_\mu(x,0,0)>0$.
\end{lemma}

\proof Set $P_{xy}=\del x\wedge\del y$ and $P_{x\mu}=\del x\wedge\del\mu$.
Using the equations of variation (\ref{eqodevar}) it is seen that
\[
P_{xy}'=-uP_{xy}+uP_{x\mu}.
\]
Two vectors which are tangent to the manifold $\wuo$ at a fixed $x$
value are
\[
\xi_1=(\rho',u',0,0,0)^T,\quad\xi_2=(0,\uy_\mu,0,1,0)^T.
\]
In the above the fact that $\uz(x,\mu,0)=0$ is implicitly used.
When applied to these vectors
\[
P_{x\mu}(\xi_1,\xi_2)=\rho',
\]
so that the equation for $P_{xy}$ becomes
\[
P_{xy}'=-uP_{xy}+\rho u^2.
\]
Since $u=\rho'/\rho$, this equation has the solution
\[
\rho(t)P_{xy}(t)=\int_{-\infty}^t\rho^2(s)u^2(s)\,ds,
\]
so that $P_{xy}(t)>0$ for all $t$.
Since for a fixed $x$ value
\[
P_{xy}(\xi_1,\xi_2)\propto\rho'\uy_\mu(x,0,0),
\]
the fact that $\rho'>0$ yields the conclusion of the lemma.
\qed

\vspace{3mm}
\begin{lemma}\label{lemsymu} For any $0<x<x_0,\,\sy_\mu(x,0,0,0)<0$.
\end{lemma}

\proof The proof follows that of the previous lemma, and will therefore
only be sketched.
Set
\[
\xi_1=(\rho',u',0,0,0)^T,\quad\xi_2=(0,\sy_\mu,0,1,0)^T.
\]
The equation for $P_{xy}$, when applied to these two vectors,
becomes
\[
P_{xy}'=-uP_{xy}+\rho u^2,
\]
which has the solution
\[
\rho(t)P_{xy}(t)=-\int_t^{\infty}\rho^2(s)u^2(s)\,ds.
\]
Since for any fixed $x$
\[
P_{xy}(\xi_1,\xi_2)\propto\rho'\sy_\mu(x,0,0,0),
\]
the conclusion immediately follows.
\qed

\vspace{3mm}
\begin{cor}\label{corfmu} For any $0<x<x_0,\,F_\mu(x,0,0)<0$.
\end{cor}

\proof The result follows immediately from the above lemmas and
equation (\ref{eqdeffmu}).
\qed

\vspace{3mm}
\begin{lemma}\label{lemuyep} $\uy_\ep(x,0,0)=0$.
\end{lemma}

\proof The proof follows that of Lemma \ref{lemuymu}, and will 
therefore again only be sketched.
Set 
\[
\xi_1=(\rho',u',0,0,0)^T,\quad\xi_2=(0,\uy_\ep,0,0,1)^T.
\]
When applied to these two vectors, the equation for $P_{xy}$ is
\[
P_{xy}'=-uP_{xy},
\]
which has the solution
\[
P_{xy}(t)=\frac{C}{\rho(t)}.
\]
Since $\lim_{t\to-\infty}P_{xy}(t)$ must be bounded, the fact that
$\rho(t)\to0$ as $t\to-\infty$ implies that the constant $C$ must
be zero.
Since for fixed $x$
\[
P_{xy}(\xi_1,\xi_2)\propto\rho'\uy_\ep(x,0,0),
\]
the conclusion of the lemma now follows.
\qed

\vspace{3mm}
The proof of the next lemma mimics those of the previous three lemmas, and
will therefore be left to the reader.
All that must be kept in mind is that the quantity $P_{xy}(t)$ must
approach zero as $t\to\infty$, as $\xi_1(t)\to0$ and the vector
$\xi_2(t)$ remains bounded as $t\to\infty$.

\begin{lemma}\label{lemsyep} $\sy_\ep(x,0,0,0)=0$.
\end{lemma}

\begin{cor}\label{corfep} $F_\ep(x,0,0)=0$.
\end{cor} 

\proof The conclusion follows immediately from the previous two
lemmas and equation (\ref{eqdeffep}).
\qed

\vspace{3mm}
As a consequence of Corollaries \ref{corfmu} and \ref{corfep}, this
section can be closed with the following theorem.

\begin{theorem}\label{thmmanintersect} There exists a smooth function 
$\mu=\mu(\ep)$ such
that when $0\le\ep\ll1$ the manifolds $\wsr$ and $\wul$ have a
nontrivial intersection.
Furthermore, the function $\mu(\ep)$ satisfies
\[
\mu(0)=\mu'(0)=0,
\]
so that
\[
\mu(\ep)=O(\ep^2)
\]
\end{theorem}

\begin{remark} Due to the symmetries present in (\ref{eqglodered}), when
$\mu=\mu(\ep)$ it is necessarily true that $\wso\cap\wume$ is also
nontrivial.
\end{remark}

\section{$\wume\cap\wsme$ transversely}
\setcounter{equation}{0}

\subsection{Flow from $\Me$ to $\{x=\eta\}$}

Assume that $\mu=\mu(\ep)=\mu_2\ep^2+O(\ep^3)$.
From the results of Section 4 it is then known that 
$\wuo\cap\wsme$ is nontrivial, as is $\wso\cap\wume$.
Furthermore, on $\Me$ there is a connection between the
critical points $(x^*(\ep),0,\pm z^*(\ep))$.
Since there exists a connection between $(0,y_-,z_-)$ and
$(0,y_+,z_+)$ in the plane $\{x=0\}$, it is clear that
when $\mu=\mu(\ep)$ there is a heteroclinic cycle in
the three-dimensional phase space.
The goal of this section is to explore the possible solution set
as $\mu$ is varied from $\mu(\ep)$.

Initially, the focus will be on the intersection of $\wuo$ and
$\wsme$; however, this is not a restriction in any way, as the
symmetry described in Proposition \ref{propsymmetry} immediately
yields similiar results for the intersection of $\wso$ and
$\wume$.
The heteroclinic wave is singular; however, it can be written in a
regular perturbation expansion for $0\le x\le x_0-\nu$.
To set notation, for $0\le x\le x_0-\nu$ write the wave
as
\begin{equation}\label{eqwaveexpand}
\begin{array}{lll}
\rho(t)&=&\rho_0(t)+O(\ep) \\
u(t)&=&u_0(t)+O(\ep) \\
\phi(t)&=&\ep\phi_1(t)+O(\ep^2).
\end{array}
\end{equation}
The following lemma gives a characterization for $\phi_1$.

\begin{lemma}\label{lemphiexpand} For $t\in(-\infty,\infty)$ the
term $\phi_1$ is given by
\[
\phi_1(t)=\be_1(t)d_1+\be_2(t)d_2,
\]
where 
\[
\begin{array}{lll}
\be_1(t)&=&-x_0 \\
\be_2(t)&=&{\ds-\frac14x_0^3(5+\tanh(\frac{x_0}{2}t))}.
\end{array}
\]
\end{lemma}

\begin{remark} Note that 
\[
\frac{\be_1(t)}{\be_2(t)}>\frac1{2x_0^2}
\]
for each $t$.
\end{remark}

\begin{remark} Although it is permissable to use $\phi_1$ when
$t\in(-\infty,T_\nu]$, where $\rho(T_\nu)=x_0-\nu$, note 
that $\be_1(t)$ and $\be_2(t)$ are well-defined for all $t$.
\end{remark}

\proof After substituting the expansion (\ref{eqwaveexpand}) into
the ODE (\ref{eqglodered}) and equating terms it is seen that
$\phi_1$ satisfies
\[
\phi_1'=-2u_0\phi_1+\om(\rho_0)-\om(x_0).
\]
Since $u_0=\rho_0'/\rho_0$, the above equation clearly has the
solution
\[
\rho_0^2(t)\phi_1(t)=\int_{-\infty}^t\rho_0^2(s)
     (\om(\rho_0(s))-\om(x_0))\,ds.
\]
Thus, after using that $\om(a)=d_1a^2+d_2a^4$ one finds that
\[
\phi_1(t)=\be_1(t)d_1+\be_2(t)d_2,
\]
where
\[
\be_i(t)=\frac1{\rho_0^2(t)}\int_{-\infty}^t\rho_0^2(s)
     (\rho_0^{2i}(s)-x_0^{2i})\,ds.
\]
Substituting the expression for $\rho_0(t)$ given in Proposition
\ref{propexactwave} into the above expression gives the 
result of the lemma.
\qed

\begin{cor}\label{corzlocation} When $\ep\neq0$ the function
$\phi(t)$ satisfies
\[
\begin{array}{lll}
\phi(T_\eta)&=&-x_0(d_1+x_0^2d_2+O(\eta))\ep+O(\ep^2) \\
\phi(T_\nu)&=&{\ds-x_0(d_1+\frac32x_0^2d_2+O(\nu))\ep+O(\ep^2)}.
\end{array}
\]
\end{cor}

\proof The result follows immediately from the expansion of the
wave given in (\ref{eqwaveexpand}) and the above lemma.
\qed

\vspace{3mm}
It is of interest to understand the behavior of $\wsme$ as it intersects
the section $\bpo$.
For fixed $\ep$ the intersection of $\wsme$ with $\bpo$ forms a
curve $y=\sy(\eta,z,\ep)$.
Setting $T_\eta$ to be such that $\rho(T_\eta)=\eta$, it will be
desirable to determine $\sy_z(\eta,\phi(T_\eta),\ep)$.
In order to find this quantity, the tangent space to the manifold
must be tracked.

In the four-dimensional phase space the manifold $\wsme$ is 
three-dimensional and can be written as the graph
\begin{equation}\label{eqwsmegraph}
\wsme=\{(x,y,z,\ep)\,:\,\eta\le x\le x_0,y=\sy(x,z,\ep)\}.
\end{equation}
Over the underlying wave, for each fixed $x$ the manifold's tangent
space is spanned by the three vectors
\begin{equation}\label{eqwsmetangent}
\begin{array}{lll}
\xi_1&=&(\rho',u',\phi',0)^T \\
\xi_2&=&(0,\sy_z,1,0)^T \\
\xi_3&=&(0,\sy_\ep,0,1)^T.
\end{array}
\end{equation}
Since $\wsme$ is a three-dimensional manifold, it will be 
necessary to use three-forms to properly track its tangent
space.
In particular, the forms 
\[
P_{xy\ep}=\del x\wedge\del y\wedge\del\ep
\]
and 
\[
P_{xz\ep}=\del x\wedge\del z\wedge\del\ep
\]
will be used.
Note that for fixed $x$
\[
P_{xy\ep}(\xi_1,\xi_2,\xi_3)=\rho'\sy_z,
\]
and in particular
\begin{equation}\label{eqsyzratio}
\sy_z(\eta,\phi(T_\eta),\ep)\propto\frac{P_{xy\ep}(T_\eta)}{\rho'(T_\eta)}.
\end{equation}
Thus, it will be desirable to compute $P_{xy\ep}$ at $x=\eta$.

After linearizing about the wave for $\ep\neq0$ the variational
equations become
\begin{equation}\label{eqvariationalepneq0}
\begin{array}{lll}
\del x'&=&u\del x+\rho\del y \\
\del y'&=&-\la'(\rho)\del x-2u\del y+2\phi\del z+(2\mu_2\ep+O(\ep^2))\del\ep \\
\del z'&=&\ep\om'(\rho)\del x-2\phi\del y-2u\del z+
           (\om(\rho)-\om(x_0)-\ep\sig)\del\ep \\
\del\ep'&=&0.
\end{array}
\end{equation}
The ODE for $P_{xy\ep}$ is then
\begin{equation}\label{eqpxyep}
P_{xy\ep}'=-uP_{xy\ep}+2\phi P_{xz\ep},
\end{equation}
which, using the fact that $u=\rho'/\rho$, has the solution
\begin{equation}\label{eqpxyepsol}
\rho(t)P_{xy\ep}(t)=(x_0-\nu)P_{xy\ep}(T_\nu)
    -2\int_t^{T_\nu}\rho(s)\phi(s)P_{xz\ep}(s)\,ds.
\end{equation}

It is clear that the evolution of $P_{xz\ep}$ must be understood in
order to understand the evolution of $P_{xy\ep}$.
Towards this end is the following proposition.

\begin{prop}\label{propxi2} When $\ep=0$,
\[
P_{xz\ep}(t)=x_0^2\frac{u_0(t)}{\rho_0(t)}.
\]
\end{prop}

\proof In order to prove the proposition, it will first be shown
that the vector $\xi_2$ is given by
\[
\xi_2(t)=(0,0,x_0^2/\rho_0^2(t),0)^T.
\]
Recall that the vector $\xi_2$ is formed by taking the
derivative of the manifold $\wsme$ with respect to $z$.
Since the flow has the symmetry $(x,y,z,t)\to(x,y,-z,t)$ when
$\ep=0$, it is clear that the first two components of $\xi_2$
must be zero.
From the variational equations the third component satisfies the
ODE
\[
\del z'=-2u_0\del z,
\]
which has the solution
\[
\del z(t)=C/\rho_0^2(t).
\]
The constant $C$ can be chosen so that 
\[
\lim_{t\to\infty}\del z(t)=1,
\]
from which arises the characterization of $\xi_2(t)$.

Since $\xi_1(t)=(\rho_0'(t),u_0'(t),0,0)$, the conclusion of
the proposition now immediately follows, as
\[
P_{xz\ep}(\xi_1,\xi_2,\xi_3)=\left|
   \begin{array}{ccc}
     \rho_0' & 0 & * \\
     0 & x_0^2/\rho_0^2 & * \\
     0 & 0 & 1
   \end{array}\right|.\quad\qed
\]

\vspace{3mm}
Recall that $T_\nu$ and $T_\eta$ are defined by
\[
\rho(T_\eta)=\eta,\quad\rho(T_\nu)=x_0-\nu.
\]
For $t\in[T_\eta,T_\nu]$ the three-form $P_{xz\ep}$ is given by
the regular perturbation expansion
\begin{equation}\label{eqpxzep}
P_{xz\ep}(t)=x_0^2\frac{u_0(t)}{\rho_0(t)}+O(\ep).
\end{equation}
Substitution of this expression into (\ref{eqpxyepsol}) along
with the expansion for $\phi$ given in Lemma \ref{lemphiexpand}
yields that
\begin{equation}\label{eqpxyepsolexpand}
\rho_0(t)P_{xy\ep}(t)=(x_0-\nu)P_{xy\ep}(T_\nu)
  -2\ep x_0^2\int_t^{T_\nu}u_0(s)(\be_1(s)d_1+\be_2(s)d_2)\,ds+O(\ep^2).
\end{equation}
In order to finish the calculation, $P_{xy\ep}(T_\nu)$ must be
determined.

As it has already been seen,
\[
P_{xy\ep}(T_\nu)=\rho'(T_\nu)\sy_z(x_0-\nu,\phi(T_\nu),\ep);
\]
thus, the quantity $\sy_z(x_0-\nu,\phi(T_\nu),\ep)$ must be
calculated.
Since $\nu$ is small, this can be determined by understanding
the variation of the linear approximation to the manifold
$\wsme$.
Since $\Me\subset\wsme$ it is necessarily true that
\begin{equation}\label{eqmesubsetwsme}
\sy(\Mx(z,\ep),z,\ep)=\My(z,\ep).
\end{equation}
Recall that Lemma \ref{lemdescribeme} states that the
manifold $\Me$ satisfies
\[
\partial_z\M_y(z,\ep)=O(\zeta^2).
\]
As such, due to (\ref{eqmesubsetwsme}) the slow manifold has no effect 
on $\wsme$, at least up to $O(\ep)$.
Therefore, in order to understand the variation of $\wsme$ it
it enough to determine how the vectors which are tangent to
$\wsme$ at $\Me$ vary with respect to $z$ and $\ep$.
Recall Proposition \ref{propevectsme}, which states that the tangent
vectors to $\wsme$ at $\Me$ are given by
\[
(-1,-\frac{v_2}{v_1},-\frac{v_3}{v_1}),
\]
where the quantities $v_i$ are defined in Proposition \ref{propevectsme}.
In order to determine $\sy_z$, it is then enough to calculate
$\partial_z(v_2/v_1)$, where these quantities are defined in
Proposition \ref{propevectsme}.
A simple calculation yields
\[
-\partial_z\frac{v_2}{v_1}=\ep\frac{\om'(x_0)}{x_0\la'(x_0)}+O(\zeta^2),
\]
from which it is seen that
\[
\sy_z(x_0-\nu,\phi(T_\nu),\ep)=\ep(\frac{\om'(x_0)}{x_0\la'(x_0)}+O(\nu))
   +O(\zeta^2).
\]
Since $\rho'(T_\nu)=O(\nu)$ one then has that
\[
P_{xy\ep}(T_\nu)=O(\nu)\ep+O(\zeta^2)
\]
Upon combining the above results, and using the fact that $\be_i(t)$ is
bounded for $i=1,2$ and for all $t$, it is seen that
\begin{equation}\label{eqpxyepfinal}
\rho_0(t)P_{xy\ep}(t)=(\al_1(t)d_1+\al_2(t)d_2+O(\nu))\ep+O(\ep^2),
\end{equation}
where
\begin{equation}\label{eqdefalphai}
\al_i(t)=-2x_0^2\int_t^\infty u_0(s)\be_i(s)\,ds
\end{equation}
for $i=1,2$.

\begin{prop}\label{propalphai} When evaluated at $t=T_\eta$,
\[
\begin{array}{lll}
\al_1(T_\eta)&=&{\ds-2x_0^3\ln\frac{\eta}{x_0}} \\
\al_2(T_\eta)&=&{\ds-2x_0^5\ln\frac{\eta}{x_0}+\frac12x_0^5
(1-\frac{\eta^2}{x_0^2})}.
\end{array}
\]
\end{prop}

\proof The expression for $\al_1(T_\eta)$ only will be proved,
as the proof for $\al_2(T_\eta)$ is similiar.
For the rest of this proof set $k=x_0/2$.

Using the expression for $\rho_0(t)$ given in Proposition
\ref{propexactwave} it is not difficult to see that
\[
\begin{array}{lll}
u_0(t)&=&\rho_0'(t)/\rho_0(t) \\
{}&=&{\ds\frac{x_0}{4}\frac{\sech^2(kt)}{1+\tanh(kt)}}.
\end{array}
\]
Since, by Lemma \ref{lemphiexpand}, $\be_1(t)=-x_0$, it is clear
that 
\[
\begin{array}{lll}
\al_1(t)&=&{\ds x_0^3\int_{kt}^\infty\frac{\sech^2(s)}{1+\tanh(s)}\,ds} \\
{}&=&{\ds x_0^3(\ln(2)-\ln(\frac{2\rho_0^2(t)}{x_0^2})}.
\end{array}
\]
The last part of the above expression comes from the fact that 
\[
1+\tanh(kt)=2\rho_0^2(t)/x_0^2.
\]
Since $\rho_0(T_\eta)=\eta$, upon simplification the expression for
$\al_1(T_\eta)$ follows.
\qed

\begin{remark}\label{remali} Note that
\[
\frac{\al_2(T_\eta)}{\al_1(T_\eta)}>x_0^2,
\]
with
\[
\frac{\al_2(T_\eta)}{\al_1(T_\eta)}=x_0^2+
   \frac14x_0^2(-\ln\frac{\eta}{x_0})^{-1}+O(\eta^2)
\]
for $0<\eta\ll1$.
\end{remark}

Now that $\al_i(T_\eta)$ has been calculated for $i=1,2$, the following
lemma has essentially been proved.
All that is now necessary is to recall (\ref{eqsyzratio}), the 
fact that $0<\rho'(T_\eta)=O(\eta)$, and the fact that 
$y_+=x_0/2$.

\begin{lemma}\label{lemsyz} The manifold $\wsme$ satisfies the 
estimate
\[
\sy_z(\eta,\phi(T_\eta),\ep)\propto\frac1{\eta^2}(\al_1d_1
+\al_2d_2+O(\nu))\ep+O(\ep^2),
\]
where
\[
\begin{array}{lll}
\al_1&=&{\ds-4x_0^2\ln\frac{\eta}{x_0}} \\
\al_2&=&{\ds-4x_0^4\ln\frac{\eta}{x_0}+x_0^4
(1-\frac{\eta^2}{x_0^2})}.
\end{array}
\]
\end{lemma}

\begin{remark} Note that $\al_1/\al_2$ satisfies the same estimate
as that given in Remark \ref{remali}.
\end{remark}

\begin{cor}\label{coruyz} The manifold $\wume$ satisfies the estimate
\[
\uy_z(\eta,-\phi(T_\eta),\ep)=\sy_z(\eta,\phi(T_\eta),\ep).
\]
\end{cor}

\proof The result follows immediately from the symmetry outlined
in Proposition \ref{propsymmetry}.
\qed

\subsection{Flow from $\{x=\eta\}$ to $\{y=y_-+\eta\}$}

Recall the transversality argument given in Section 4.
It was shown how the manifolds $\wuo$ and $\wsme$ vary with
respect to $\mu$ along the wave when $\ep=0$.
Due to the smoothness of the flow, these estimates persist for
$\ep$ and $\mu$ sufficiently small.
Furthermore, as a by-product of the symmetry discussed in Proposition
\ref{propsymmetry}, the manner in which $\wso$ and $\wume$ vary
with respect to $\mu$ is also understood.

Specifically, the following is known.
Set
\begin{equation}\label{eqwumeandwso}
\begin{array}{rll}
\wume\cap\{x=\eta\}&=&\{(y,z,\mu,\ep)\,:\,y=\uy(z,\mu,\ep)\} \\
\wso\cap\{x=\eta\}&=&\{(y,z,\mu,\ep)\,:\,y=\sy(\mu,\ep),
z=\sz(\mu,\ep)\}.
\end{array}
\end{equation}
As a consequence of Proposition \ref{propsymmetry} and Lemmas
\ref{lemuymu} and \ref{lemsymu} it can be concluded that
\begin{equation}\label{eqwumewsovary}
\sy_\mu(0,0)<0,\quad\uy_\mu(0,0,0)>0;
\end{equation}
furthermore, these estimates hold for $\ep$ and $\mu$ sufficiently
small.
When $\mu=\mu(\ep)$ the manifolds intersect, so that
\begin{equation}\label{eqwumewsointersect}
\sy(\mu,\ep)=\uy(\sz(\mu,\ep),\mu,\ep).
\end{equation}
Thus, when $\ep$ is fixed, by the estimate (\ref{eqwumewsovary}) 
for $\mu>\mu(\ep)$ the manifold $\wso$ is below $\wume$, while for 
$\mu<\mu(\ep)$ the configuration is reversed.

\begin{figure}[htb]
\epsfysize=8.5in
\epsffile{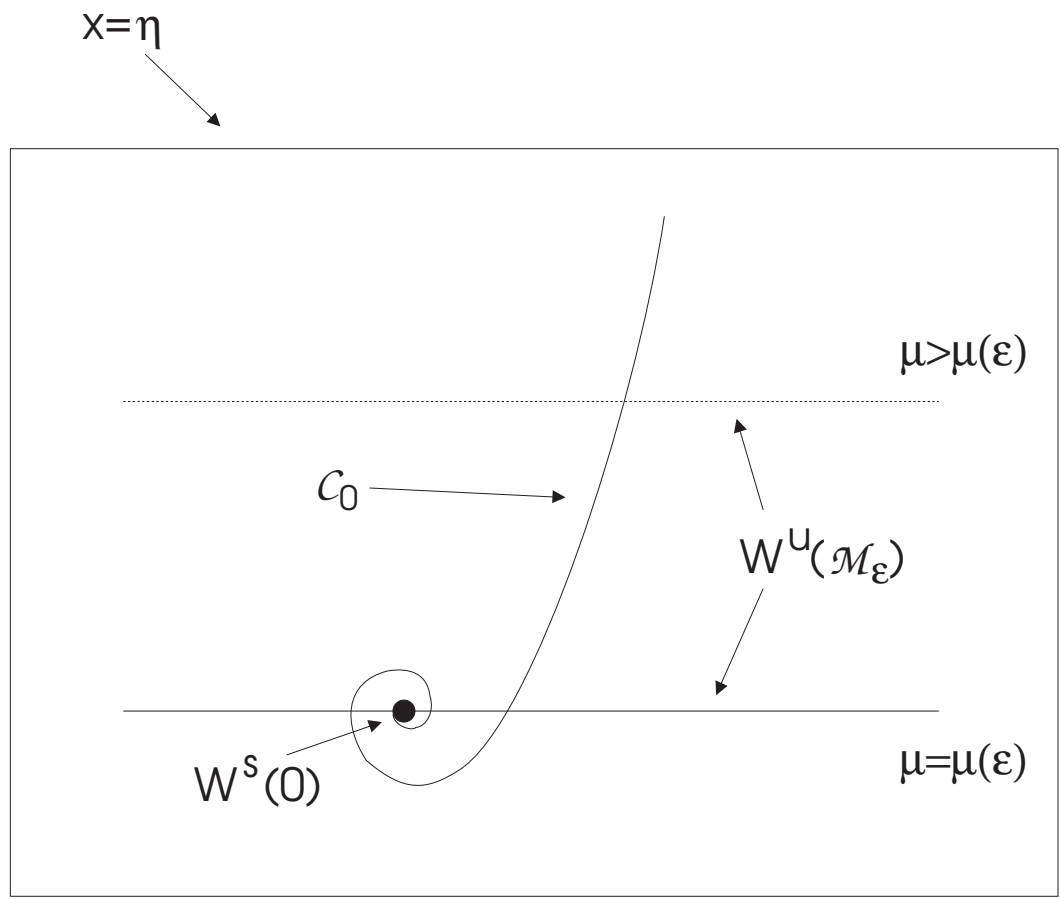}
\caption{Existence of dark solitary waves\label{fig_wume}}
\end{figure}

Recall Lemma \ref{lemspirals}, which gives a description of
the curves $\co\cdot t_0^\pm$ and ${\cal C}_{\pm\infty}\cdot t_{\pm\infty}$.
Specifically, it is of interest here to contemplate the nature of
$\co\cdot\tmo$ and $\cmi\cdot\tmi$, both of which are contained
in $\bmo$ and centered upon
\begin{equation}\label{eqdefpso}
\pls=\wso\cap\bmo.
\end{equation}
The set
\begin{equation}\label{eqdefcmo}
\cmo=\cmi\cdot\tmi\cup\pls\cup\co\cdot\tmo
\end{equation}
defineds a curve in $\bmo$; in fact, for $\ep\om(x_0)\neq0$ it
is a two-armed logarithmic spiral centered upon $\pls$.
When $\mu=\mu(\ep),\,\wume$ intersects $\cmo$ at $\pls$; indeed,
the intersection is transverse.
Thus, as $\mu$ is varied from $\mu(\ep),\,\wume$ continues to
intersect $\cmo$; however, the intersection no longer occurs at
$\pls$.
This implies that $\wume$ intersects either $\cmi\cdot\tmi$ or
$\co\cdot\tmo$.
From (\ref{eqwumewsovary}) and Lemma \ref{lemspirals} it can be
concluded that if $|\mu-\mu(\ep)|$ is $O(\ep^n)$ for $n\ge3$ but
is not exponentially small in $\ep$, then 
\begin{equation}\label{eqwumeintersect}
\begin{array}{lll}
\mu>\mu(\ep)&\Longrightarrow&\wume\cap\co\cdot\tmo\neq\emptyset \\
\mu<\mu(\ep)&\Longrightarrow&\wume\cap\cmi\cdot\tmi\neq\emptyset
\end{array}
\end{equation}
(see Figure \ref{fig_wume}).
Furthermore, there exists only one point in each of the
intersections. 
If $0<|\mu-\mu(\ep)|\le O(e^{-c/\ep})$, then there exists a
finite number of points in both $\wume\cap\co\cdot\tmo$ and
$\wume\cap\cmi\cdot\tmi$, with the number of points increasing
to infinity as $\mu\to\mu(\ep)$.

For the rest of this subsection it will be assumed that
$|\mu-\mu(\ep)|$ is not exponentially small in $\ep$.
This is done to clarify the following arguments.
In addition, it will be assumed that $\mu\ge\mu(\ep)$;
however, this is not a necessary restriction and the below
arguments can be modified to draw conclusions if $\mu<\mu(\ep)$.

Since $\mu>\mu(\ep)$, by (\ref{eqwumeintersect}) $\wume\cap\co\cdot\tmo$
is nonempty, which immediately implies that $\wume\cap\co$ is
nonempty.
Due to symmetry outlined in Proposition \ref{propsymmetry} this
yields not only that $\wsme\cap\co$ is nonempty, but
\[
\wume\cap\co=\wsme\cap\co\subset\wume\cap\wsme.
\]
Thus, there is a connection between $(x^*(\ep),0,z^*(\ep))$ and
$(x^*(\ep),0,-z^*(\ep))$ which passes ``near'' (to be made more
precise later) the plane $\{x=0\}$.
The goal of the rest of this subsection is to understand the
nature in which $\wume$ and $\wsme$ intersect at $\co$.

Now that the orientation of the manifolds $\wume$ and $\wsme$ is
understood as they intersect the plane $\{x=\eta\}$, it is desirable
to understand their behavior under the flow as they pass near
the critical points $(0,y_-,z_-)$ and $(0,y_+,z_+)$,
respectively.
The goal of this subsection is to show that under a suitable
restriction the passage near these critical points does not
effect their orientation.
It will be sufficient to track the manifold $\wume$, as the
symmetry present in (\ref{eqglodered}) allows one to then draw an
immediate conclusion regarding the behavior of $\wsme$.

First, suppose that $\mu$ is such that
\begin{equation}\label{eqmurestrict}
0<O(\ep^{n+1})\le|\mu-\mu(\ep)|\le O(\ep^n)
\end{equation}
for some $n\ge3$.
As it will be seen, this restriction on $\mu$ will guarantee that
the spiralling action near the critical points on $\{x=0\}$ will be 
unseen by $\wume$ as it passes near $\{x=0\}$.
Since $n\ge2$, the perturbation of $\mu$ given in (\ref{eqmurestrict})
will not effect the asymptotic estimates given in Lemma \ref{lemsyz}
and Corollary \ref{coruyz}.
Using the asymptotics for $\uy$, it will be desirous to look at
the image of the curve
\begin{equation}\label{equycurve}
x=\eta,\quad y=y_0+\ep cz,
\end{equation}
where $|z|\le\eta,\,y_0=O(\ep^n),$ and
\[
\sgn(c)=\sgn(\uy_z).
\]
The point $y_0$ describes the difference between $\mu$ and 
$\mu(\ep)$, and by supposition is therefore positive.

It will be advantageous to look at the flow near the critical point
in polar coordinates.
Upon setting
\[
y=r\sin\th,\quad z=r\cos\th
\]
the curve (\ref{equycurve}) can be parametrically described as
\begin{equation}\label{equypolarcurve}
\begin{array}{lll}
r_0(s)&=&\sqrt{y_0^2+2c\ep s+(1+\ep^2c^2)s^2}\\
\th_0(s)&=&{\ds\mbox{Tan}^{-1}(\frac{y_0+\ep cs}{s})},
\end{array}
\end{equation}
where $|s|\le\eta$.
The linear flow near $(0,y_-,z_-)$ satisfies the ODE
\begin{equation}\label{eqzerolinear}
\begin{array}{lll}
x'&=&-ax \\
r'&=&2ar \\
\th'&=&\ep b,
\end{array}
\end{equation}
where
\[
\begin{array}{lll}
a&=&\sqrt{-\la(0)}+O(\ep^2)\\
b&=&{\ds\frac{\om(x_0)}{\sqrt{-\la(0)}}+O(\ep)}.
\end{array}
\]
Thus, upon using (\ref{equypolarcurve}) as an initial condition 
to (\ref{eqzerolinear}) it is seen that the behavior of $\uy_z$
is well approximated by the solution formula
\begin{equation}\label{equypolarflow}
\begin{array}{lll}
x(t)&=&\eta e^{-at}\\
r(t)&=&r_0(s)e^{2at}\\
\th(t)&=&\th_0(s)+\ep bt.
\end{array}
\end{equation}

It is now of interest to determine the nature of (\ref{equypolarcurve})
as the flow forces it to intersect the cylinder $\{r=\eta\}$.
The time of flight, $t_f$, is defined by $r(t_f)=\eta$.
Using (\ref{equypolarflow}), this yields that
\[
t_f=\frac{1}{2a}\ln\frac{\eta}{r_0(s)}.
\]
Thus, under the linear flow (\ref{equypolarflow}) there is the
mapping from $\{x=\eta\}$ to $\{r=\eta\}$ given by
\begin{equation}\label{equymap}
\left(\begin{array}{c}
  \eta \\ r_0(s) \\ \th_0(s)
      \end{array}\right)\to
\left(\begin{array}{c}
  \sqrt{\eta}\sqrt{r_0(s)} \\ \eta \\
  {\ds\th_0(s)+\ep\frac{b}{2a}\ln\frac{\eta}{r_0(s)}}
      \end{array}\right)
\end{equation}
for $|s|\le\eta$.
Set $x(s)$ to be the first component and $\th(s)$ to be the third
component of the vector on the right hand side of the above 
equation.
Given the asymptotic expansion for $\uy_z$, it is then of interest
to perform an expansion of $x(s)$ and $\th(s)$ around $s=0$.
This step is taken because it is only of interest to determine
the behavior of the curve near where $\wume$ intersects $\co\cdot\tmo$.
First, a Taylor expansion yields that
\[
\begin{array}{lll}
r_0(s)&=&y_0+\ep cs+O(w^2) \\
\th_0(s)&=&{\ds\frac{\pi}2-\frac1{y_0}s+O(w^2)},
\end{array}
\]
where $w=\ep s/y_0$ and it has been used that $y_0>0$.
Plugging this expansion into (\ref{equymap}) and performing
another Taylor expansion then gives that
\begin{equation}\label{equypolartaylor}
\begin{array}{lll}
x(s)&=&{\ds\sqrt{\eta}(\sqrt{y_0}+\frac{\ep c}{2\sqrt{y_0}}s
         +O(w^2))} \\
\th(s)&=&{\ds\frac{\pi}2+\ep\frac{b}{2a}\ln\frac{\eta}{y_0}
         -\frac1{y_0}(1+\ep^2\frac{ac}{2b})s+O(w^2)}.
\end{array}
\end{equation}
In rectangular coordinates the right hand side of (\ref{equymap})
is given by $x(s),\,y(s)=\eta\sin\th(s),$ and $z(s)=\eta\cos\th(s)$.
Using the expansion (\ref{equypolartaylor}) it is then seen that 
\begin{equation}\label{equyrecttaylor}
\begin{array}{llllll}
x(0)&=&\sqrt{\eta}\sqrt{y_0},\quad&
x'(0)&=&{\ds\sqrt{\eta}\frac{\ep c}{2\sqrt{y_0}}} \\
y(0)&=&\eta\sin\th(0),\quad&
y'(0)&=&{\ds-\eta(\frac1{y_0}(1+\ep^2\frac{ac}{2b}))\cos\th(0)} \\
z(0)&=&\eta\cos\th(0),\quad&
z'(0)&=&{\ds\eta(\frac1{y_0}(1+\ep^2\frac{ac}{2b}))\sin\th(0)}.
\end{array}
\end{equation}

It is clear that the curve given by the right hand side of 
(\ref{equymap}) can be parameterized by $z$ for $s$ sufficiently
near zero.
Using the expansion coefficients given in (\ref{equyrecttaylor}) it
is then seen that $x=x(z)$ satisfies the relation
\begin{equation}\label{equyderiv}
{\ds\frac{dx}{dz}=\frac{\ep c\sqrt{y_0}}{2\sqrt{\eta} 
(1+\ep^2\frac{ac}{2b}))\sin\th(0)}}.
\end{equation}
Since
\[
\th(0)=\frac{\pi}2+\ep\frac{b}{2a}\ln\frac{\eta}{y_0},
\]
this means that for $y_0=O(\ep^n)$,
\[
\th(0)=\frac{\pi}2+O(n\ep\ln\frac1{\ep}).
\]
Substituting this estimate into (\ref{equyderiv}) and applying 
(\ref{equyrecttaylor}) yields the following lemma.

\begin{lemma}\label{lemuyzb0} Suppose that $\mu-\mu(\ep)=O(\ep^n)$ for
some $n\ge2$.
The intersection of $\wume$ with the plane $\{y=y_-+\eta\}$ is a 
parametric curve given by $(x,z)=(x(s),z(s))$.
The curve satisfies the estimates
\[
\begin{array}{ll}
a.\quad&x(0)=O(\ep^{n/2}) \\
b.\quad&{\ds\left.\frac{dx}{dz}\right|_{s=0}=O(\ep^{(n+2)/2})} \\
c.\quad&{\ds\sgn(\frac{dx}{dz})=
           \sgn(\uy_z(\eta,-\phi(T_\eta),\mu(\ep),\ep))}.
\end{array}
\]
\end{lemma}

\begin{remark} If $\wume$ intersects the plane $\{y=y_--\eta\}$, then
the conclusion of the lemma holds, with part c. being changed to
\[
{\ds\sgn(\frac{dx}{dz})=
          -\sgn(\uy_z(\eta,-\phi(T_\eta),\mu(\ep),\ep))}.
\]
\end{remark}

As an immediate consequence of the symmetry of the ODE one gets
the following corollary.

\begin{cor}\label{corsyzb0} Suppose that $\mu-\mu(\ep)=O(\ep^n)$ for
some $n\ge2$.
The intersection of $\wsme$ with the plane $\{y=y_+-\eta\}$ is a 
parametric curve given by $(x,z)=(x(s),z(s))$.
The curve satisfies the estimates
\[
\begin{array}{ll}
a.\quad&x(0)=O(\ep^{n/2}) \\
b.\quad&{\ds\left.\frac{dx}{dz}\right|_{s=0}=O(\ep^{(n+2)/2})} \\
c.\quad&{\ds\sgn(\frac{dx}{dz})=
           \sgn(\sy_z(\eta,\phi(T_\eta),\mu(\ep),\ep))}.
\end{array}
\]
\end{cor} 

\begin{remark} If $\wume$ intersects the plane $\{y=y_++\eta\}$, then
the conclusion of the lemma holds, with part c. being changed to
\[
{\ds\sgn(\frac{dx}{dz})=
          -\sgn(\uy_z(\eta,\phi(T_\eta),\mu(\ep),\ep))}.
\]
\end{remark}

\subsection{Flow from $\{y=y_-+\eta\}$ to $\{y=0\}$}

Now that the passage of the manifold $\wume$ near the critical point
$(0,y_-,z_-)$ is understood, it is necessary to understand its
behavior under the flow near $\{x=0\}$.
The previous lemma states that the manifold is $O(\ep^{n/2})$ near
$\{x=0\}$ when it intersects the plane $\{y=y_-+\eta\}$.
Since $x'=xy$ and $y<0$ in the region of interest, it is clear
that manifold stays within $O(\ep^{n/2})$ of $\{x=0\}$ until 
it intersects the plane $\{y=0\}$.
Thus, it is expected that the flow on $\{x=0\}$ dominates the
behavior of the manifold.
The goal of this subsection is to make this intuition rigorous.

After the manifold $\wume$ has intersected the plane $\{y=y_-+\eta\}$
it can be written as the graph
\begin{equation}\label{eqdefux}
\wume=\{(x,y,z)\,:\,x=\ux(y,z,\ep),y_-+\eta\le y\le y_+-\eta,|z|\le\eta\}.
\end{equation}
The previous lemma gives an expression for $\ux_z$ along the wave when
$y=y_-+\eta$.
The goal is to calculate $\ux_z(0,0,\ep)$.
In order to accomplish this, it will be necessary to once again use
differential forms.

The vector to be tracked has the initial condition
\[
\xi=(\ux_z(y_-+\eta,z,\ep),0,1)^T,
\]
where the $z$-coordinate is taken to be such that the vector is
over the underlying solution.
The variational equations are
\begin{equation}\label{eqvarnear0}
\begin{array}{lll}
\del x'&=&u\del x+\rho\del y \\
\del y'&=&-\la'(\rho)\del x-2u\del y+2\phi\del z \\
\del z'&=&\ep\om'(\rho)\del x-2\phi\del y-2u\del z.
\end{array}
\end{equation}
Projectivize the equations by setting 
\begin{equation}\label{eqdefproject}
a=\del x/\del z,\quad b=\del y/\del z.
\end{equation}
It is not difficult to check that $\del z(\xi)=O(1)$ for 
$y_-+\eta\le y\le 0$, so that the above quantities are
well-defined in the region of interest.
The equations for $a$ and $b$ are given by
\begin{equation}\label{eqprojectodes}
\begin{array}{lll}
a'&=&(-u+2\phi b)a+\rho b-\ep\om'(\rho)a^2 \\
b'&=&-\la'(\rho)a+2\phi(1+b^2)-\ep\om'(\rho)ab,
\end{array}
\end{equation}
with the initial conditions being
\[
\begin{array}{l}
a(0)=\ux_z(u\cdot0,\phi\cdot0,\ep) \\
b(0)=0.
\end{array}
\]

Without loss of generality, assume that the $z$-coordinate
of $\wu(0)\cap\bmo,\,-\phi(T_\eta),$ is such that $-\phi(T_\eta)>0$.
By Corollary \ref{corzlocation}, this implies that 
\begin{equation}\label{eqfirstd1d2restrict}
d_1+x_0^2d_2>0.
\end{equation}
Let $(\rho,u,\phi)\cdot t\in\wume\cap\wsme$ represent the orbit
connecting $(x^*(\ep),0,\pm z^*(\ep))$, and let it be normalized so
that $u\cdot0=y_-+\eta$.  
It is known that this orbit intersects $\co$, as $\wume\cap\co\cdot\tmo$
is nontrivial.
Therefore, there is a $T_0>0$ such that $u\cdot T_0=\phi\cdot T_0=0$.
In addition, there exists a $T_1<0$ such that $\rho\cdot T_1=\eta$.

The claim is that $\phi\cdot t>0$ for $t\in[T_1,T_0]$.
To see that this is true, consider the following argument.
The $z$-coordinate of $\wume\cap\co\cdot\tmo,\,\phi\cdot T_1$, is
within $O(\ep^3),\,n\ge3$, of $-\phi(T_\eta)$; therefore, since
$-\phi(T_\eta)>0$ is $O(\ep),\,\phi\cdot T_1$ is also positive.
Following the argument of the previous subsection, it is not 
difficult to see that $\phi\cdot t>0$ for $t\in[T_1,0]$.
Since $\phi\cdot0>0$, as a consequence of Proposition \ref{propco}
it is necessarily true that $\phi\cdot t>0$ for $t\in[0,T_0]$
(recall that, by Lemma \ref{lemuyzb0}, $\rho\cdot0=O(\ep^{n/2})$).

As a consequence of Corollary \ref{coruyz} and Lemma \ref{lemuyzb0} it
is known that 
\[
\begin{array}{l}
\sgn(\ux_z(u\cdot 0,\phi\cdot 0,\ep))=\sgn(\al_1d_1+\al_2d_2) \\
\ux_z(u\cdot 0,\phi\cdot 0,\ep)=O(\ep^{(n+2)/2}).
\end{array}
\]
In what follows, assume that
\begin{equation}\label{eqsecondd1d2restrict}
\al_1d_1+\al_2d_2>0.
\end{equation}
Since $\al_1/\al_2$ is close to $1/x_0^2$ (for an exact description
see Lemma \ref{lemsyz}), this is not that much more a restriction than
that already imposed by (\ref{eqfirstd1d2restrict}).

Under this hypothesis, it is then being assumed that $a(0)>0$, with
$a(0)=O(\ep^{(n+2)/2})$.
It is of interest to calculate $a(T_0)$, as
\begin{equation}\label{eqat0}
\ux_z(0,0,\ep)\propto a(T_0).
\end{equation}
Towards this end is the following lemma.

\begin{lemma}\label{lemagrowthestimate} $a(t)\le O(\ep^{n/2})$ for
$t\in[0,T_0]$.
\end{lemma}

\proof First, since $\rho'=\rho u$,
\[
\rho(t)=\rho(0)e^{\int_0^tu(s)\,ds},
\]
so that $\rho(t)=O(\rho(0))=O(\ep^{n/2})$ for $t\in[0,T_0]$.
In addition, since $\del z(\xi)$ is $O(1)$ for $y_-+\eta\le y\le0$, and
$\del x(\xi)$ and $\del y(\xi)$ satisfy the linear ODE (\ref{eqvarnear0}),
it is necessarily true that $a$ and $b$ are $O(1)$ for $t\in[0,T_0]$.
These two facts will be used extensively in what follows.

Set
\[
\begin{array}{lll}
f(t)&=&{\ds e^{\int_0^tu(s)-2\phi(s)b(s)\,ds}} \\
    &=&{\ds\frac{\rho(t)}{\rho(0)}e^{-2\int_0^t\phi(s)b(s)\,ds}}.
\end{array}
\]
Since $b(t)=O(1)$ and $\rho(t)=O(\rho(0))$, it is clear that $f(t)=O(1)$ for 
$t\in[0,T_0]$.
Solving (\ref{eqprojectodes}), it is seen that $a(t)$ is given by
the integral formula
\begin{equation}\label{eqasol}
f(t)a(t)=a(0)+\int_0^tf(s)\rho(s)b(s)\,ds-
             \ep\int_0^tf(s)\om'(\rho(s))a^2(s)\,ds.
\end{equation}
Using the estimates on $f$ and $\rho$, and assuming that $a$ and
$b$ are $O(1)$, it can then be concluded that
\[
f(t)a(t)=a(0)+O(\ep^{n/2})\cdot t+O(\ep^{(n+2)/2})\cdot t.
\]
Since $a(0)=O(\ep^{(n+2)/2})$, the conclusion follows.
\qed

\vspace{3mm}
It will be desirable to revise the above estimate.
In order to do so, the following lemma is needed.

\begin{lemma}\label{lembgrowthestimate} $b(t)=O(\ep)$ for
$t\in[0,T_0]$.
\end{lemma}

\proof Since $b(0)=0$, the function $b(t)$ satisfies the integral
equation
\begin{equation}\label{eqbsolution}
b(t)=2\int_0^t\phi(s)(1+b^2(s))\,ds-\int_0^t(\la'(\rho(s))a(s)
+\ep\om'(\rho(s))a(s)b(s))\,ds.
\end{equation}
Assuming that $b(t)=O(1)$, which was justified in the proof
of the above lemma, using the conclusion of the above
lemma, and using the fact that $\phi(t)=O(\ep)$, one gets
that
\[
b(t)=O(\ep)\cdot t+O(\ep^{n})\cdot t.
\]
Since $T_0=O(1)$, the conclusion now follows.
\qed

\vspace{3mm}
A more careful examination of (\ref{eqbsolution}) yields the
following result for $b(t)$, which in turn can be used to improve upon
the estimate made upon $a(t)$ in Lemma \ref{lemagrowthestimate}.
Using the estimates provided in the above two lemmas, it is clear
that 
\[
b'=2\phi(1+b^2)+O(\ep^n),
\]
which can be solved to get
\[
b(t)=\tan(\int_0^t2\phi(s)\,ds)+O(\ep^n)\cdot t.
\]
Substituting this expression into (\ref{eqasol}), and again using
the above two lemmas, yields
\[
f(t)a(t)=a(0)+\int_0^tf(s)\rho(s)\tan(\int_0^s2\phi(r)\,dr)\,ds
+O(\ep^{3n/2})\cdot t.
\]
Note that the second term on the right-hand side of the above
equation is $O(\ep^{(n+2)/2})$, as is $a(0)$.
Furthermore, this term is positive, as $\phi(t)>0$ for 
$t\in[0,T_0]$.
Since, by supposition, $a(0)>0$, it is now clear that
\[
f(T_0)a(T_0)>a(0),
\]
and furthermore, $a(T_0)=O(a(0))$.

Using (\ref{eqat0}), the proof of the following lemma is now
complete.
While the lemma was proved only for the case of both $\phi\cdot0>0$ and
$a(0)>0$, it can easily be modified in the event that both quantities
are negative.

\begin{lemma}\label{lemuxzy=0} Suppose that $0<\mu-\mu(\ep)=O(\ep^n)$
for some $n\ge3$.
Further suppose that
\[
\sgn(\al_1d_1+\al_2d_2)=\sgn(d_1+x_0^2d_2),
\]
where the $\al_i$'s are defined in Lemma \ref{lemsyz}.
Then 
\[
\begin{array}{l}
\sgn(\ux_z(0,0,\ep))=\sgn(\al_1d_1+\al_2d_2) \\
\ux_z(0,0,\ep)=O(\ep^{(n+2)/2}).
\end{array}
\]
\end{lemma}

By the symmetries present in (\ref{eqglodered}) one arrives at
the following corollary.

\begin{cor}\label{corsxzy=0} Let the hypotheses of Lemma \ref{lemuxzy=0}
be satisfied.
Then
\[
\sgn(\sx_z(0,0,\ep))=-\sgn(\ux_z(0,0,\ep)).
\]
\end{cor}

Using the time of flight estimates given in Subsection 5.2, one finally
gets the following theorem.

\begin{theorem}\label{thmwumewsmeintersect} Suppose that 
$0<\mu-\mu(\ep)=O(\ep^n)$ for some $n\ge3$.
Further suppose that the parameters $d_1$ and $d_2$ are chosen
so that 
\[
\sgn(\al_1d_1+\al_2d_2)=\sgn(d_1+x_0^2d_2),
\]
or
where the $\al_i$'s are defined in Lemma \ref{lemsyz}.
Then the manifolds $\wume$ and $\wsme$ intersect transversely 
at $\{y=0\}$,
with the transversality being $O(\ep^{(n+2)/2})$.
Furthermore, the time the resultant trajectory spends in the region
$0\le x\le\eta$ is given by
\[
T_f=O(2n\ln\frac1{\ep}).
\]
\end{theorem}

\subsection{Completion of argument}

Using the notation of the previous section, near the invariant plane
$\{x=0\}$ write
\[
\begin{array}{lll}
\wume=\{(x,y,z,\ep)\,:\,x=\ux(y,z,\ep),y_-+\eta\le y\le y_+-\eta\} \\
\wsme=\{(x,y,z,\ep)\,:\,x=\sx(y,z,\ep),y_-+\eta\le y\le y_+-\eta\}.
\end{array}
\]
By Theorem \ref{thmwumewsmeintersect} it is known that if
$0<\mu-\mu(\ep)=O(\ep^n)$ for $n\ge3$ and if the pair $(d_1,d_2)$ satisfy
the hypotheses of Lemma \ref{lemuxzy=0}, then there exists a
$(\tilde{x}(\ep),0,0)\in\co,\,\tilde{x}(\ep)=O(\ep^{n/2})$, such that
\[
\begin{array}{l}
\ux(0,0,\ep)=\sx(0,0,\ep) \\
\sgn((\ux_z-\sx_z)(0,0,\ep))=\sgn(\al_1d_1+\al_2d_2) \\
(\ux_z-\sx_z)(0,0,\ep)=O(\ep^{(n+2)/2}).
\end{array}
\]
Since the manifolds intersect transversely at $(\tilde{x}(\ep),0,0)$, 
there is an orbit, 
$(\rho(t),u(t),\phi(t))$, which connects the two critical
points $(x^*(\ep),0,\pm z^*(\ep))\in\Me$.
Let this trajectory by translated so that $(\rho(0),u(0),\phi(0))=
(\tilde{x}(\ep),0,0)\in\co$.

Since the manifolds $\wume$ and $\wsme$ intersect transversely
at $\{y=0\}$, they intersect transversely everywhere along
the orbit.
The next goal is to understand the transversality as $\wume$ intersects
$\bpxo$.
Define
\begin{equation}\label{eqdeftangentvectors}
\begin{array}{lll}
\xi_1&=&(\rho'(0),u'(0),\phi'(0))^T \\
\xi_2&=&(\ux_z(\tilde{x}(\ep),0,1)^T \\
\xi_3&=&(\sx_z(\tilde{x}(\ep),0,1)^T.
\end{array}
\end{equation}
It is clear that $\xi_2\cdot t\in T\wume$ and $\xi_3\cdot t\in T\wsme$
for all $t$, and that $\xi_1\cdot t\in T\wume\cap T\wsme$ for all
$t$.
The three-form $P_{xyz}$ will be used to gain an understanding
as to how the flow carries these vectors.
Using (\ref{eqvarnear0}), the three-form $P_{xyz}$ satisfies
Abel's formula
\[
P_{xyz}'=-3uP_{xyz},
\]
which, using the fact that $u=\rho'/\rho$, has the solution
\begin{equation}\label{eqpxyzsolvetemp}
P_{xyz}(t)=\frac{\rho^3(0)}{\rho^3(t)}P_{xyz}(0).
\end{equation}
When applied to the tangent vectors $\xi_i$ defined in
(\ref{eqdeftangentvectors}),
\begin{equation}\label{eqpxyz0}
P_{xyz}(0)=(\la(0)+O(\ep^2))(\ux_z-\sx_z)(0,0,\ep);
\end{equation}
thus, after substituting into (\ref{eqpxyzsolvetemp}) one gets
\begin{equation}\label{eqpxyzsolve}
P_{xyz}(t)=\frac{\rho^3(0)}{\rho^3(t)}
     (\la(0)+O(\ep^2))(\ux_z-\sx_z)(0,0,\ep).
\end{equation}

Now define $\tilde{T}_\nu$ such that $\rho(\tilde{T}_\nu)=x_0-\nu$.
As they intersect the section $\{x=x_0-\nu\}$ the manifolds 
$\wume$ and $\wsme$ are given by the curves
\[
\begin{array}{lll}
\wume\cap\bpxo&=&\{(y,z,\ep)\,:\,y=\uy(z,\ep)\} \\
\wsme\cap\bpxo&=&\{(y,z,\ep)\,:\,y=\sy(z,\ep)\}.
\end{array}
\]
In addition, there exists a $\tilde{z}(\ep)$ such that
\[
\uy(\tilde{z}(\ep),\ep)=\sy(\tilde{z}(\ep),\ep).
\]

In order to understand the transversality of the manifolds
at $\{x=x_0-\nu\}$ it is necessary to compute
$(\uy_z-\sy_z)(\tilde{z}(\ep),\ep)$.
Towards this end, set
\[
\begin{array}{lll}
\tilde{\xi}_2&=&(0,\uy_z(\tilde{z}(\ep),\ep),1)^T \\
\tilde{\xi}_3&=&(0,\sy_z(\tilde{z}(\ep),\ep),1)^T.
\end{array}
\]
It is clear that
\begin{equation}\label{eqpxyzpropto}
\begin{array}{lll}
P_{xyz}(\tilde{T}_\nu)&\propto&
    P_{xyz}(\xi_1\cdot\tilde{T}_\nu,\tilde{\xi}_2,\tilde{\xi}_3) \\
{}&=&\rho'(\tilde{T}_\nu)(\uy_z-\sy_z)(\tilde{z}(\ep),\ep).
\end{array}
\end{equation}
Since $\rho'(\tilde{T}_\nu)>0$ is $O(\nu)$, upon substituting
(\ref{eqpxyzsolve}) into (\ref{eqpxyzpropto}) one gets
\begin{equation}\label{eqtransverseatx0}
\begin{array}{lll}
(\uy_z-\sy_z)(\tilde{z}(\ep),\ep)&\propto&
    \rho^3(0)\la(0)(\ux_z-\sx_z)(0,0,\ep) \\
{}&=&A\ep^{2n+1}+O(\ep^{2n+2}),
\end{array}
\end{equation}
where $\sgn(A)=-\sgn(\al_1d_1+\al_2d_2)$ (recall that $\la(0)<0$).
The argument for the following lemma is now complete.

\begin{lemma}\label{lemtransverseatx0} Let $0<\mu-\mu(\ep)=O(\ep^n)$ 
for some $n\ge3$.
There exists a $\tilde{z}(\ep)$, with $|\tilde{z}(\ep)-\phi(T_\nu)|=
O(\ep^{n-1})$, such that
\[
\begin{array}{ll}
1.\quad&\uy(\tilde{z}(\ep),\ep)=\sy(\tilde{z}(\ep),\ep) \\
2.\quad&\sgn((\uy_z-\sy_z)(\tilde{z}(\ep),\ep))=
       -\sgn(\al_1d_1+\al_2d_2) \\
3.\quad&(\uy_z-\sy_z)(\tilde{z}(\ep),\ep)=O(\ep^{2n+1}).
\end{array}
\]
\end{lemma}

\begin{remark} To paraphrase, the manifold $\wume$ intersects $\wsme$
transversely at the section $\{x=x_0-\nu\}$, with the transversality being
$O(\ep^{2n+1})$ for $0<\mu-\mu(\ep)=O(\ep^n)$.
\end{remark}

\section{Existence of solitary waves}
\setcounter{equation}{0}

\subsection{Existence of bright solitary waves}

A bright solitary wave is characterized by
\[
\lim_{t\to\pm\infty}x(t)=0;
\]
hence, in order to have such a wave it is necessary that
$\wuo\cap\wso$ be nontrivial.
Due to the symmetry described in Proposition \ref{propsymmetry},
it is enough to show that $\wuo\cap\cxo\cdot\txo$, and hence
$\wuo\cap\cxo$.

Recall Lemma \ref{lemimagecxo}, which states that $\cxo\cdot\txo$ is
$C^1-O(e^{-c/\ep})$ close to $\wsme$ for points $p\in\cxo$ which take
$O(1/\ep)$ time to exit $\bxo$.
Under a time reversal, the flow on $\Me$ for $z=O(\ep)$ satisfies
\[
z'=\ep^2\sig+O(\ep^3).
\]
Therefore, $\cxo\cdot\txo$ will be close to $\wsme$ in a region where
$z=O(\ep)$ with $\sgn(z)=\sgn(\sig)$.

\begin{figure}[htb]
\epsfysize=8.5in
\epsffile{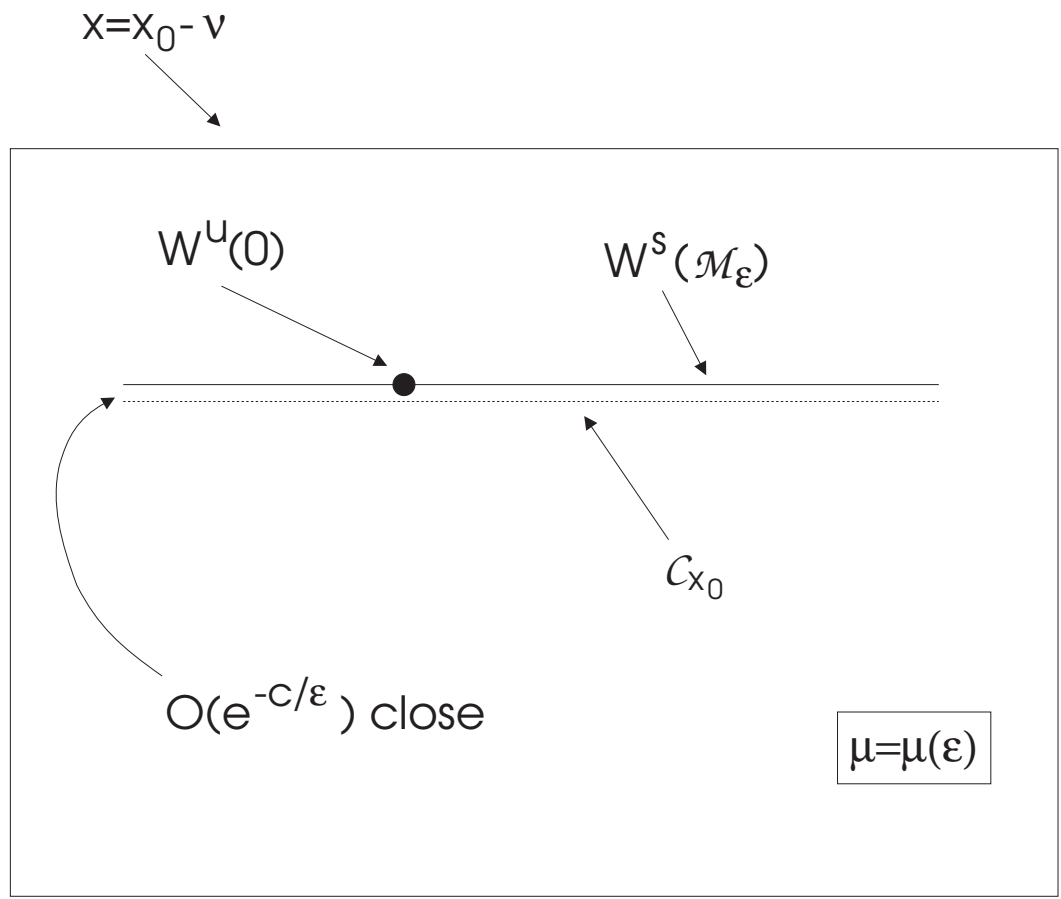}
\caption{Existence of bright solitary waves\label{fig_wsme}}
\end{figure}

\begin{figure}[htb]
\epsfysize=8.5in
\epsffile{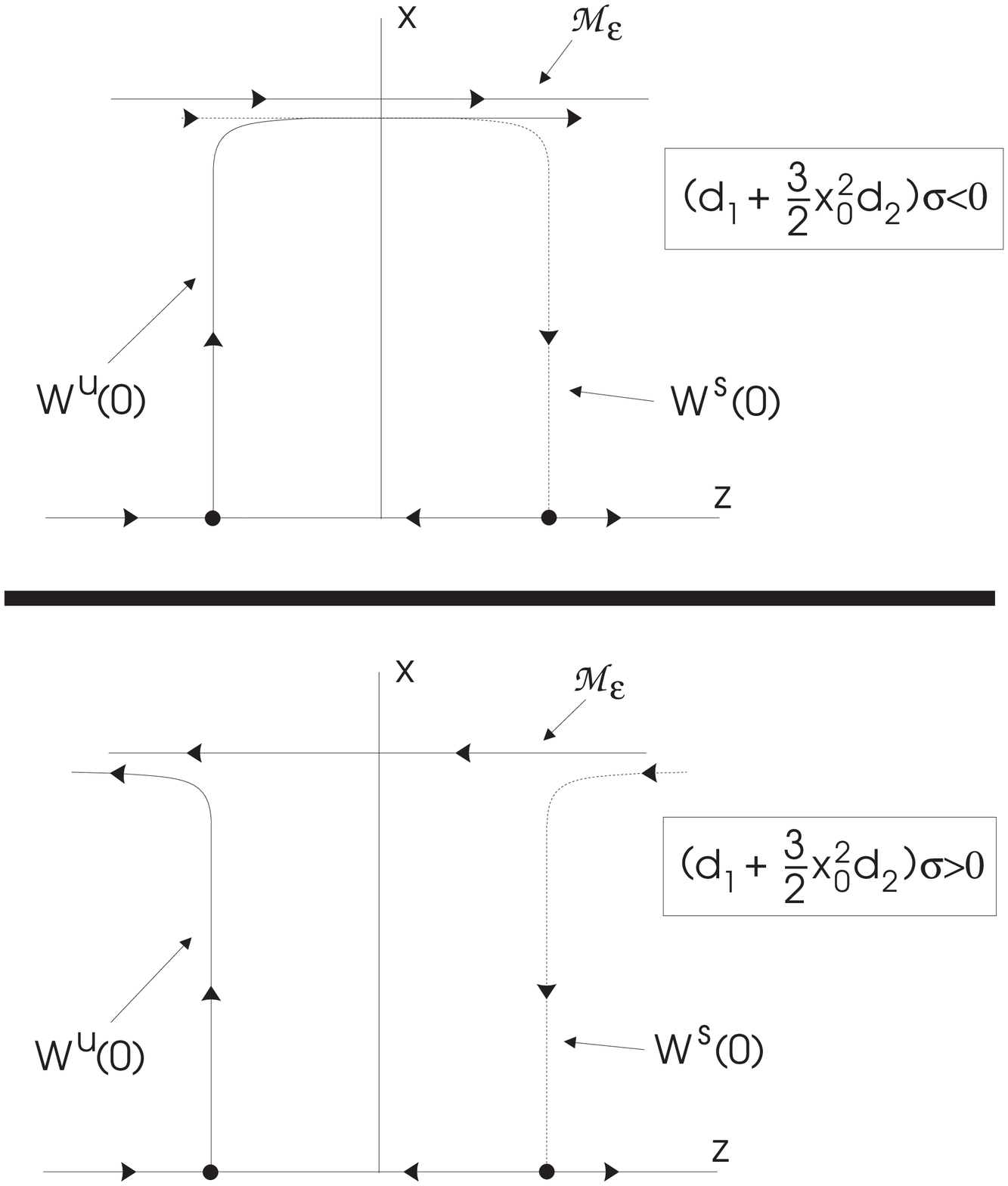}
\caption{Projection of flow on $\{y=0\}$\label{fig_projflow}}
\end{figure}

Now recall Corollary \ref{corzlocation}, which states that when
$\mu=\mu(\ep)$ that $\phi(T_\nu)$, the $z$-coordinate of $\wuo\cap\bpxo$,
is given by
\[
\phi(T_\nu)=-x_0(d_1+\frac32x_0^2d_2+O(\nu))\ep+O(\ep^2).
\]
Thus, by the comments of the previous paragraph, if 
\[
(d_1+\frac32x_0^2d_2)\sig<0,
\]
then $\wuo\cap\wsme$ is $C^1$-$O(e^{-c/\ep})$ close to the curve
$\cxo\cdot\txo$ (see Figures \ref{fig_wsme} and \ref{fig_projflow}).
Using the facts that $\cxo\cdot\txo$ is $C^1$-$O(e^{-c/\ep})$ close
to $\wsme$ and that $\wuo$ intersects $\wsme$ transversely yield
the following lemma.

\begin{figure}[htb]
\epsfysize=8.5in
\epsffile{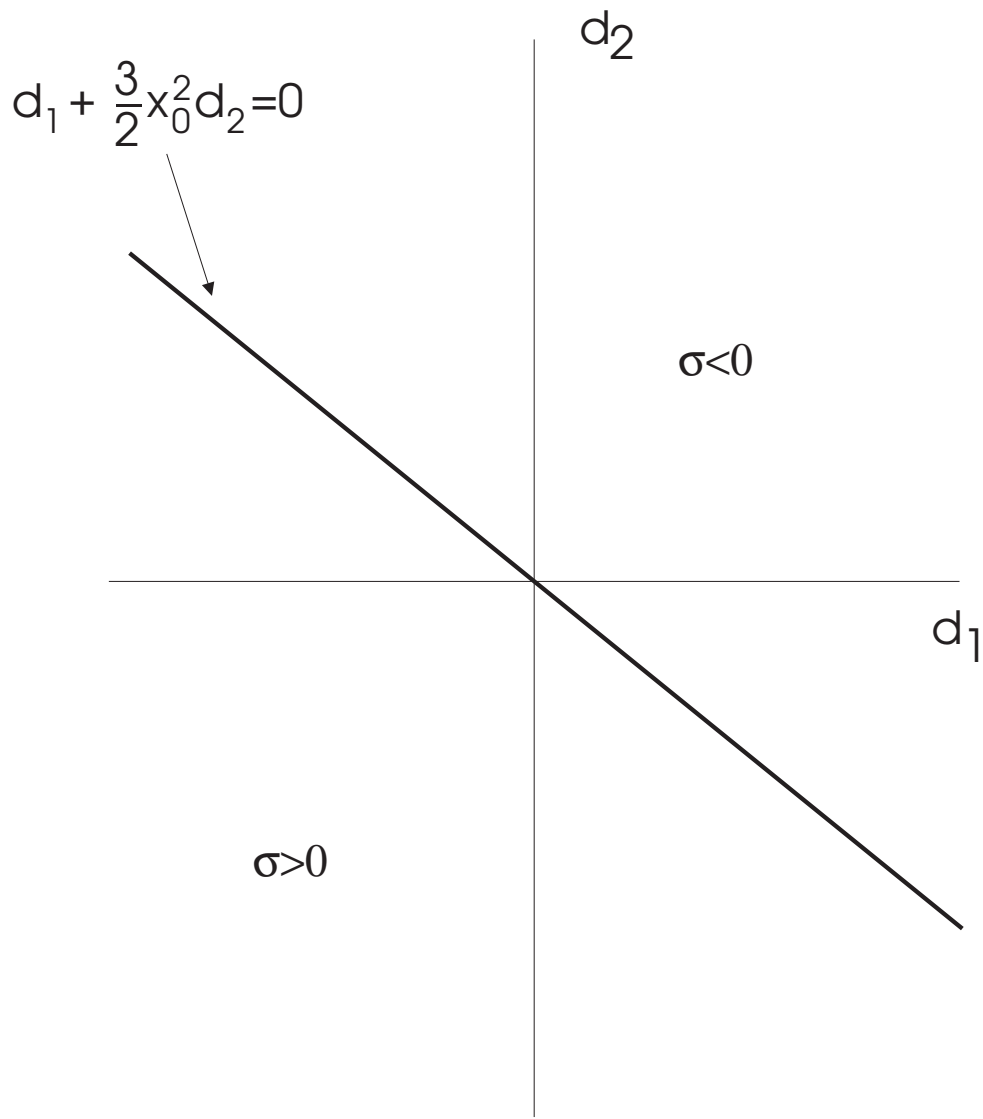}
\caption{Parameter regime for bright solitary waves\label{fig_hompar}}
\end{figure}

\begin{lemma}\label{lem1brightwave} Suppose that
\[
(d_1+\frac32x_0^2d_2)\sig<0
\]
(see Figure \ref{fig_hompar}).
There exists a $\mu_h(\ep)<\mu(\ep)$, with $\mu(\ep)-\mu_h(\ep)=
O(e^{-c/\ep})$, such that when $\mu=\mu_h(\ep)$, then $\wuo\cap\wso$
is nontrivial.
\end{lemma}

\begin{remark} The fact that $\mu(\ep)-\mu_h(\ep)>0$ is a 
consequence of Lemmas \ref{lemuymu} and \ref{lemsymu}.
\end{remark}

The solution described in the above lemma can be thought of as a
1-pulse solution.
It is characterized by the fact $\wuo$ intersects the plane
$\{y=0\}$ at exactly one point.
An $N$-pulse solution is characterized by both 
$\wuo\cap\wso$ being nontrivial and $\wuo$ intersecting $\{y=0\}$ at
$N$ distinct points.
Given the 1-pulse solution, it is natural to inquire as to the
existence of $N$-pulses.
Fortunately, this question has been studied in the work of
Kapitula and Maier-Paape \cite{kapitula:sdo96}.
In order to quote the results stated in that paper, the next
lemma must first be proved.

\begin{lemma}\label{lemwoutransverse} Let $0<\ep\ll1$ be 
fixed.
Let $z_h(\mu)$ represent the $z$-coordinate of $\wuo\cap\{y=0\}$.
Then
\[
\begin{array}{ll}
1.\quad&z_h(\mu_h(\ep))=0 \\
2.\quad&{\ds\frac{d}{d\mu}z_h(\mu_h(\ep))\neq0}.
\end{array}
\]
\end{lemma}

\proof The first part of the conclusion follows immediately from
the fact that when $\mu=\mu_h(\ep),\,\wuo\cap\cxo\neq\emptyset$.

For fixed $\ep$ the set $\wuo\cap\bpxo$ yields a curve parameterized
by $\mu$, i.e.,
\[
\wuo\cap\bpxo=\{(y,z,\mu)\,:\,y=y^u(\mu), z=z^u(\mu)\}.
\]
Due to Lemma \ref{lemuymu} it is known that
\[
\frac{d}{d\mu}y^u(0)>0,
\]
so that 
\begin{equation}\label{eqyumu}
\frac{d}{d\mu}y^u(\mu_h(\ep))>0,
\end{equation}
as $\mu_h(\ep)=O(\mu(\ep))=O(\ep^2)$.

Let $p=(x_0-\nu,y^u(\mu_h(\ep)),z^u(\mu_h(\ep)))\in\wuo\cap\bpxo$, 
and let $T_h>0$ be such that $p\cdot T_h\in\cxo$.
Since $z'=-\ep^2\sig+O(\ep^3)$ and $z^u(\mu_h(\ep))=O(\ep)$, it
is necessarily true that $T_h=O(1/\ep)$.
Therefore, as a consequence of the Exchange Lemma with Exponentially
Small Error \cite{jones:tim96}, the transversality described
by (\ref{eqyumu}) gets transferred into a transversality condition
in the slow direction.
Since the slow direction near $\Me$ is described by $z$, this means
that 
\[
\frac{d}{d\mu}z_h(\mu_h(\ep))\neq0.\quad\qed
\]

\vspace{3mm}
With the above lemma in hand, it is now possible to state Theorem 
1.7 in Kapitula and Maier-Paape \cite{kapitula:sdo96}.

\begin{theorem}\label{thmNbrightwaves} Let $0<\ep\ll1$, and suppose that
\[
(d_1+\frac32x_0^2d_2)\sig<0.
\]
For each $N\ge2$ there
exists a bi-infinite sequence $\{\mu^N_k\}$ such that when
$\mu=\mu^N_k$ there is an $N$-pulse solution to (\ref{eqglodered}).
The $N$-pulse is such that $x(t)$ is even in $t$.
Furthermore,
\[
|\mu^N_k-\mu_h(\ep)|=O(e^{-c|k|/\ep})
\]
as $|k|\to\infty$.
\end{theorem}

\begin{remark} The estimate on $|\mu^N_k-\mu_h(\ep)|$ is not
explicitly provided in Theorem 1.7 of Kapitula and Maier-Paape;
however, it is implicit in the proof of that theorem.
\end{remark}

\begin{remark} Although one can discuss the existence of $N$-pulses
which are odd in $t$, this will not be done here.
For a more complete description of the dynamical behavior for
$\mu$ near $\mu_h(\ep)$, the interested reader should consult
\cite{kapitula:sdo96}.
\end{remark}

\subsection{Existence of dark solitary waves}

A dark solitary wave is characterized by
\[
\lim_{t\to\pm\infty}(x,y,z)\cdot t\in\Me;
\]
hence, in order to have such a wave it is necessary that both
$\wume\cap\wsme\neq\emptyset$ and there exist critical points
in $\Me$.
By Proposition \ref{asssig_1}, the existence of the critical points
is guaranteed if
\begin{equation}\label{eqd1d2cond1}
(d_1+2x_0^2d_2)\sig<0.
\end{equation}
By Lemma \ref{lemtransverseatx0}, $\wume\cap\wsme\neq\emptyset$ if
\begin{equation}\label{eqd1d2cond2}
\begin{array}{l}
0<\mu-\mu(\ep)=O(\ep^n),\quad n\ge3 \\
(d_1+x_0^2d_2)(\al_1d_1+\al_2d_2)>0,
\end{array}
\end{equation}
where, by Lemma \ref{lemsyz}, the coefficients $\al_i$ satisfy
\[
\frac{\al_2}{\al_1}=x_0^2+\frac14x_0^2(-\ln\frac{\eta}{x_0})^{-1}
+O(\eta^2).
\]
Furthermore, the manifolds intersect transversely, with the 
transversality being $O(\ep^{2n+1})$ (Lemma \ref{lemtransverseatx0}).

An $N$-circuit solution is a dark solitary wave whose trajectory
passes near $\{x=0\}\,N$ times.
If (\ref{eqd1d2cond1}) and (\ref{eqd1d2cond2}) are satisfied, then
for $0<\ep\ll1$ there is a 1-circuit solution.
The goal is to show that there exist $N$-circuits for $2\le N<N(\ep)$,
where $N(\ep)\to\infty$ as $\ep\to0$.
This will be accomplished through the use of the Exchange Lemma with
Exponentially Small Error (ELESE) (Jones et al \cite{jones:tim96}).

In order for the ELESE to apply, it must first be shown that
$\wume\cap\wsme$ transversely.
This task was accomplished in the series of lemmas leading to Lemma
\ref{lemtransverseatx0}.
The ELESE then applies to all those points in
\[
\cpu=\wume\cap\bpxo
\]
which take $O(1/\ep)$ time to exit $\bxo$.

Set
\[
\cps=\wsme\bpxo.
\]
The $z$-coordinate of $\cpu\cap\cps$, say $\hat{z}$, is given by
\[
\hat{z}=-x_0(d_1+\frac32x_0^2d_2+O(\nu))\ep+O(\ep^2),
\]
i.e., $|\hat{z}-\phi(T_\nu)|=O(\ep^2)$.
This is due to the assumption that $\mu-\mu(\ep)=O(\ep^n)$ for
some $n\ge3$.
Set
\[
\cmu=\wume\cap\bmxo,\quad\cms=\wsme\cap\bmxo.
\]
Due to the symmetry inherent in the ODE, not only is
$\cmu\cap\cms\neq\emptyset$, the $z$-coordinate of the
intersection is given by $-\hat{z}$.
Now, recall that for $|z|=O(\ep)$ the flow on $\Me$ is given by
\[
z'=-\ep^2\sig+O(\ep^3)
\]
(equation \ref{eqmeflowzsmall}).
Thus, if
\begin{equation}\label{eqd1d2cond3}
(d_1+\frac32x_0^2d_2)\sig<0
\end{equation}
the flow on $\Me$ is from $\hat{z}$ to $-\hat{z}$; otherwise, it
is in the opposite direction.

Assume that (\ref{eqd1d2cond3}) holds.
Since $z'=O(\ep^2)$, it takes a time of $O(1/\ep)$ for trajectories
to traverse from $\hat{z}$ to $-\hat{z}$.
Let $\cpu\cdot T^+_u$ represent the intersection
of the curve $\cpu$ with the section $\bmxo$ as it is carried by the flow 
generated by the ODE (\ref{eqglodered}).
By the ELESE the curve $\cpu\cdot T^+_u$ will be $C^1$-$O(e^{-c/\ep})$
close to $\cmu$ at $-\hat{z}$.
Since $\cmu\cap\cms$ transversely at $-\hat{z}$, with the order
of transversality being $O(\ep^{2n+1})$, this implies that 
$\cpu\cdot T^+_u\cap\cms$ transversely exponentially close to $-\hat{z}$.
In other words, upon passing through $\bxo$ the manifold
$\wume$ becomes oriented so that it once again intersects
$\wsme$ transversely.
Given the existence of a 1-circuit solution, the existence of a
2-circuit has now been proven.

The above argument can be repeated to show the existence of $N$-circuit
solutions for each $N\ge2$.
Given an $\ep>0$, the maximal $N$, say $N(\ep)$, will be such that
$N(\ep)\to\infty$ as $\ep\to0$.
This is due to the fact that upon passage through $\bxo$, the curve
$\cpu\cdot T^+_u$ is  $C^1$-$O(e^{-c/\ep})$
close to $\cmu$ at $-\hat{z}$.
The following theorem has now been proven.

\begin{figure}[htb]
\epsfysize=8.5in
\epsffile{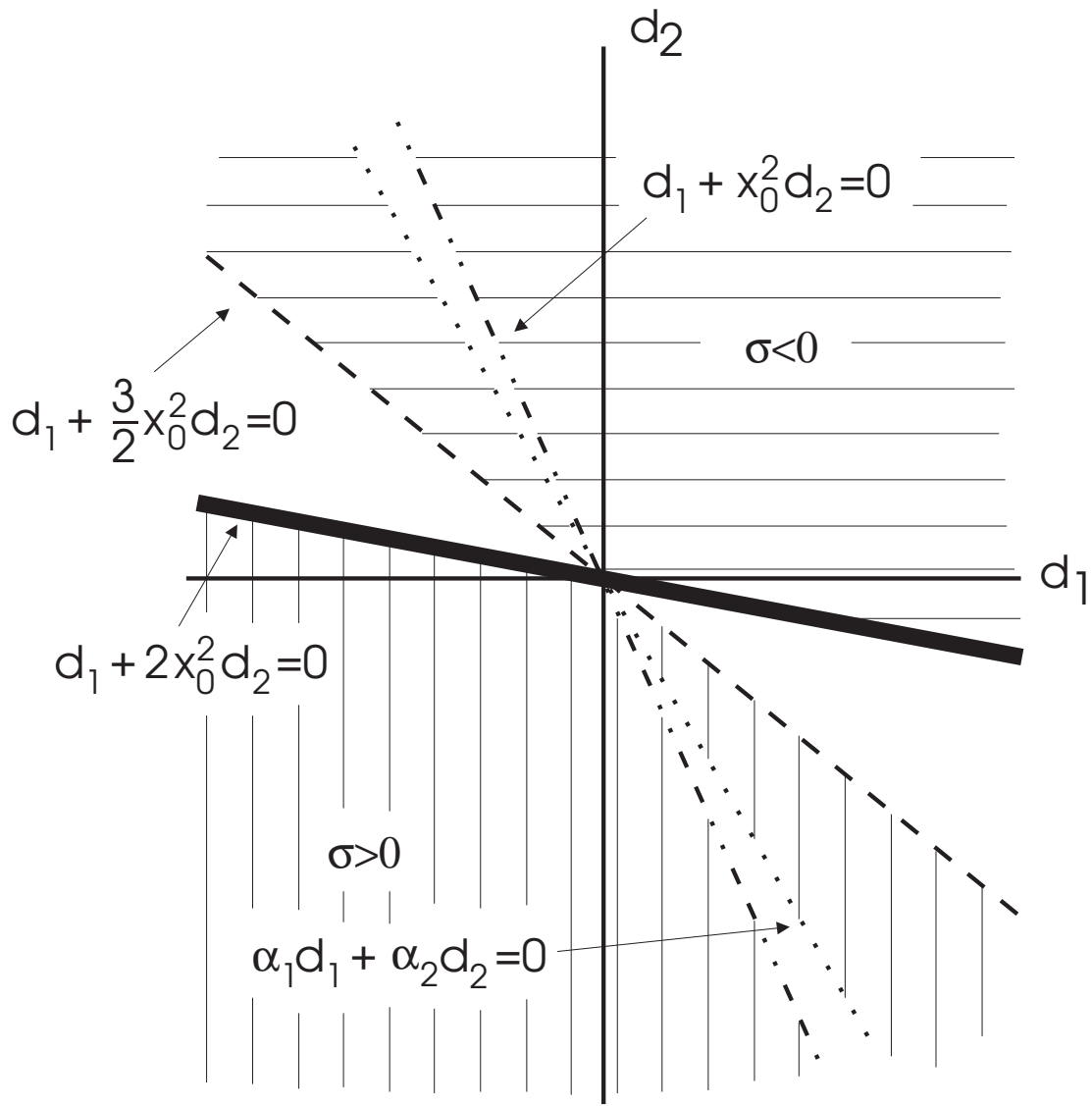}
\caption{Parameter regime for dark solitary waves\label{fig_n_cirpar}}
\end{figure}

\begin{theorem}\label{thmNdarkwaves} Suppose that
$0<\ep\ll1$, and suppose that (\ref{eqd1d2cond1}), (\ref{eqd1d2cond2}),
and (\ref{eqd1d2cond3}) hold (see Figure \ref{fig_n_cirpar}).
Then there exists an $N(\ep)>1$, with $N(\ep)\to\infty$ as
$\ep\to0$, such that $N$-circuit solutions exist for $1\le N<N(\ep)$.
\end{theorem}

\bibliography{papers}
\bibliographystyle{plain}

\newpage

\newpage

\newpage

\newpage

\newpage

\newpage

\newpage


\end{document}